\documentclass[10pt,aps,prl,twocolumn,superscriptaddress]{revtex4-1}
\usepackage{amsmath}
\usepackage{amsfonts}
\usepackage{amssymb}
\usepackage{graphicx}
\usepackage[english]{babel}

\usepackage{setspace}
\renewcommand{\l}{\left(}
\renewcommand{\r}{\right)}

\renewcommand{\H}{\hat{\mathcal{H}}}

\renewcommand{\a}{\hat{a}}

\newcommand{\ad}{\hat{a}^\dagger}

\begin{document}

%\preprint{}

\title{Microscopy of the interacting Harper-Hofstadter model in the few-body limit}


\author{M. Eric Tai}
\author{Alexander Lukin}
\author{Matthew Rispoli}
\author{Robert Schittko}
\author{Tim Menke}
\author{Dan Borgnia}
\affiliation{Department of Physics, Harvard University, Cambridge, Massachusetts 02138, USA}
\author{Philipp M. Preiss}
\affiliation{Physikalisches Institut, Universit\"at Heidelberg, 69120 Heidelberg, Germany}
\author{Fabian Grusdt}
\author{Adam M. Kaufman}
\author{Markus Greiner}
\email{greiner@physics.harvard.edu}
\affiliation{Department of Physics, Harvard University, Cambridge, Massachusetts 02138, USA}
%\thanks{}


\date{\today}
\newcommand{\ket}[1]{|#1\rangle}

\begin{abstract}
The interplay of magnetic fields and interacting particles can lead to exotic phases of matter exhibiting topological order and high degrees of spatial entanglement~\cite{chen_local_2010}. While these phases were discovered in a solid-state setting~\cite{tsui_FQH_1982,laughlin_FQH_1983}, recent techniques have enabled the realization of gauge fields in systems of ultracold neutral atoms~\cite{madison_dynamical_1998,lignier_dynamical_2007,jaksch_creation_2003,kolovsky_creating_2011,madison_vortex_2000,abo-shaeer_observation_2001,lin_synthetic_2009,aidelsburger_experimental_2011,an_direct_2016,kolkowitz_spin-orbit_2016}, offering a new experimental paradigm for studying these novel states of matter. This complementary platform holds promise for exploring exotic physics in fractional quantum Hall systems due to the microscopic manipulation and precision possible in cold atom systems~\cite{sorensen_fractional_2005,hafezi_fractional_2007,gemelke_rotating_2010}. However, these experiments thus far have mostly explored the regime of weak interactions~\cite{schweikhard_rapidly_2004,zwierlein_vortices_2005,jotzu_experimental_2014,
stuhl_visualizing_2015,mancini_observation_2015,aidelsburger_realization_2013,miyake_realizing_2013}. Here, we show how strong interactions can modify the propagation of particles in a $2\times N$, real-space ladder governed by the Harper-Hofstadter model~\cite{harper_1955,hofstadter_1976}. We observe inter-particle interactions affect the populating of chiral bands, giving rise to chiral dynamics whose multi-particle correlations indicate both bound and free-particle character. The novel form of interaction-induced chirality observed in these experiments demonstrates the essential ingredients for future investigations of highly entangled topological phases of many-body systems.
\end{abstract}

%\pacs{123456}
%\keywords{}

\maketitle

% body of paper here - Use proper section commands~~\cite
% References should be done using the ~\cite, \ref, and \label commands
%\section{Introduction}

\makeatletter %only needed in preamble
\renewcommand\small{\@setfontsize\small{8pt}{9.5}}
\makeatother

\begin{figure}[htb]
    \centering
    \includegraphics[width=\columnwidth]{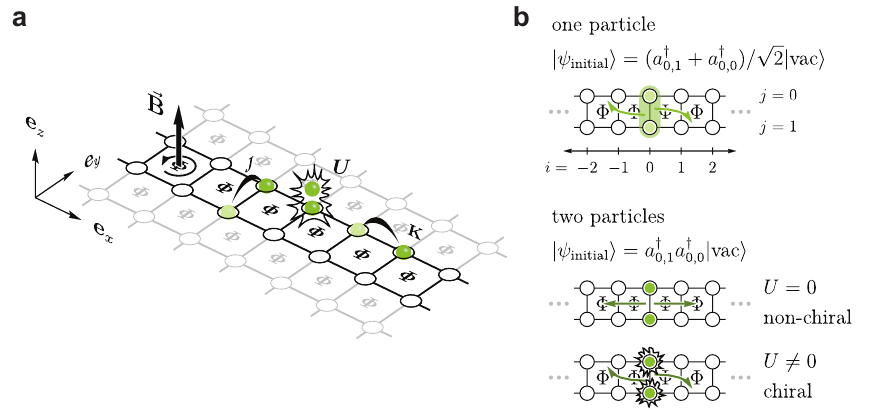}
    \caption{\label{fig:ladder_schematic}
    \textbf{a}, From a 2D lattice, we isolate a $2 \times N$ ladder region in which we study the Harper-Hofstadter model. Nearest neighbor lattice sites in the $x$ and $y$ direction are coupled by complex- and real-valued tunnelings with magnitudes $K$ and $J$, respectively, realizing an artificial gauge field with constant flux per unit cell $\Phi$. When multiple atoms occupy the same lattice site, they experience a pairwise interaction shift $U$.
	\textbf{b}, We first study the motion of a single particle delocalized over two sites of a given rung. Due to the coupling of motion in the $x$ and $y$ directions induced by the gauge field, chiral dynamics emerge where rightward (leftward) motion is correlated with a bias towards the lower (upper) leg of the ladder. However, a pair of non-interacting particles initialized onto opposite sides of a single rung does not exhibit chirality even in the presence of a gauge field. The addition of interactions breaks the symmetry between particles going to the left and to the right, thereby reintroducing chiral motion.}
\end{figure}

The Harper-Hofstadter Hamiltonian is a model for describing lattice systems in the presence of a gauge field. When the system is confined to a ladder or strip geometry, two  bands of opposite chirality emerge and give rise to chiral edge modes~\cite{huegel_chiral_2014}. Edge modes have been a subject of increasing interest in the condensed matter and quantum information communities~\cite{Nayak2008} because of their topologically-protected properties, which arise due to the absence of backscattering in the presence of chirality~\cite{Hasan2010}. They have been studied in photonic systems~\cite{Hafezi2013}, and in atomic lattice systems with synthetic dimensions using macroscopic observables~\cite{stuhl_visualizing_2015,mancini_observation_2015}. While synthetic dimensions provide an elegant solution to realize well-controlled, finite systems~\cite{Celi2014}, we introduce a tunable, real-space, finite lattice that can be extended in straightforward ways to studies in arbitrary  geometries (Figure~\ref{fig:ladder_schematic}a). We create a finite ladder in which we perform microscopic studies of the edge modes at the single-particle level: we prepare a well-defined initial state to load these edge modes and allow the state to evolve under the Hamiltonian to expose the chiral nature of the dynamics (Figure~\ref{fig:ladder_schematic}b). An imbalanced decomposition of the initial state into the two chiral bands results in shearing in the particle propagation, i.e. a particle is more likely to occupy a specific leg of the ladder and which leg is dependent on its velocity. This shearing indicates a coupling of the dynamics along the leg ($x$) and rung ($y$) directions of the ladder that is reminiscent of a Lorentz force.  

In the presence of interactions, the dynamics of even two particles is substantially modified from the expectation obtained from a single-particle picture. The interactions modify the eigenspectrum such that states of both scattering and bound nature emerge, the population of which governs the observed dynamics. Importantly, the interactions provide an avenue through which states of a certain chirality can be preferentially populated, giving rise to chiral trajectories in situations where such chirality would be absent for vanishing interactions, even in the presence of a gauge field (Figure~\ref{fig:ladder_schematic}b). By exploiting the toolset of quantum gas microscopy~\cite{bakr_quantum_2009,sherson_single_2010} --- spatially resolved correlations and single-particle observables --- we experimentally identify the mechanisms through which interactions produce these chiral dynamics. Importantly, while the measurements performed here are all non-equilibrium, the excellent agreement between theory and experiment, in the presence of interactions, paves the way for equilibrium measurements of chiral ground states~\cite{Dhar2012,sorensen_fractional_2005,hafezi_fractional_2007, Roushan2016}.

\begin{figure}[!htb]
    \centering
    \includegraphics[width=\columnwidth]{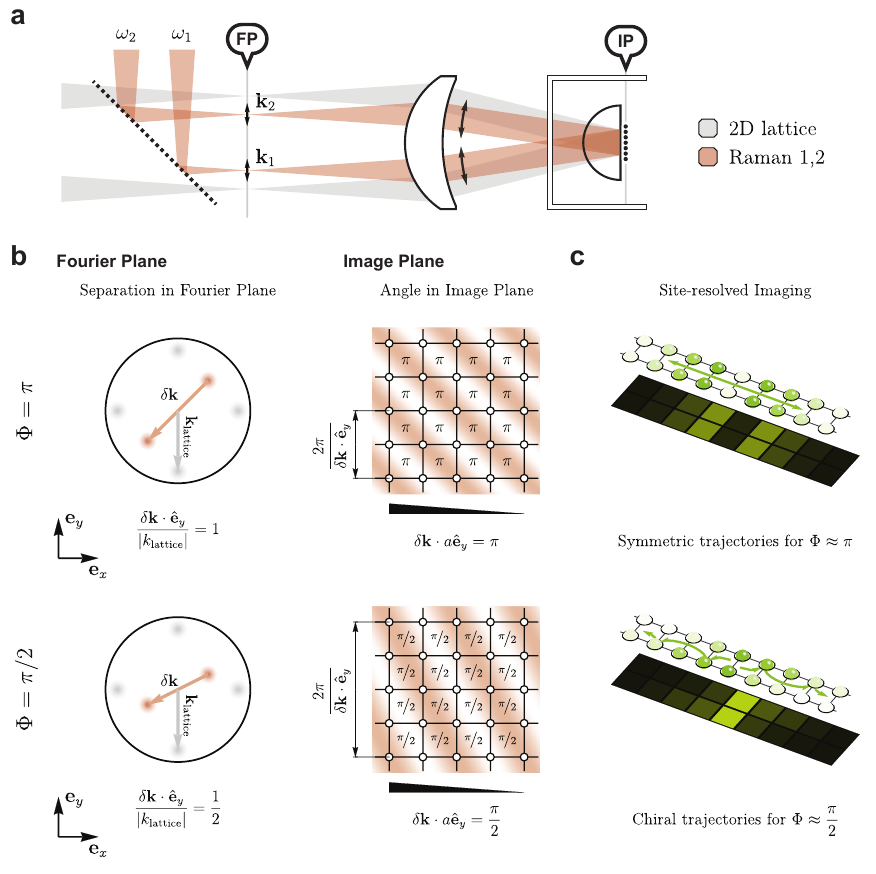}
    \caption{\label{fig:experiment_schematic}
    Using a quantum gas microscope, we can realize artificial gauge fields of variable strength by projecting a running lattice with variable spacing through the same objective used to project the static $2\text{D}$ lattice and image the atoms. 
    \textbf{a}, A pair of beams (in gray) interfere in the image plane of the microscope, creating one axis of the $2\text{D}$ optical lattice. A pair of Raman beams with wavevectors $\mathbf k_1$ and $\mathbf k_2$ and slightly different frequencies $\omega_1$ and $\omega_2$ (in brown) also interfere in the image plane, creating a running lattice. We can adjust the periodicity and orientation of the Raman lattice by moving the position of the beams in the Fourier plane of the microscope.
    \textbf{b}, Top (bottom) row: example realizing flux $\Phi = \pi$ ($\Phi = \pi /2$) in the system. Fourier plane of the microscope: gray (brown) disks correspond to the beams used to create the $2\text{D}$ (Raman) lattice with resulting wavevector $k_\text{lattice}$ ($\delta \mathbf k \equiv \mathbf k_1 - \mathbf k_2$). The ratio of the $y$ component of the Raman lattice wave vector and the lattice wave vector determines the flux in the system~\cite{aidelsburger_experimental_2011,miyake_realizing_2013}. Resulting image plane structure: The $2\text{D}$ lattice is represented by the black grid. The brown shading represents the spatial intensity distribution of the running Raman lattice at one instance in time. The black triangle indicates the potential gradient imposed by a physical magnetic field, which is used to detune lattice sites along the $x$ direction.
	\textbf{c}, Resulting dynamics for example cases of $\Phi \approx \pi$ and $\Phi \approx \pi/2$ flux. The top diagram for each example illustrates the time dynamics of the center of mass for particles traveling to the left and to the right from the initial rung. The bottom diagram is the measured density distribution following time evolution, exhibiting asymmetric propagation for a flux $\Phi \approx \pi/2$.}
\end{figure}

Our experiments begin with a two-dimensional Bose-Einstein condensate of Rubidium-87 atoms. These atoms sit at the focus of a high-resolution imaging system through which we project a square optical lattice with spacing $a = 680\, \text{nm}$. By collecting atomic fluorescence, we detect the parity of site occupations on the individual site level. On this platform, we realize the Harper-Hofstadter Hamiltonian,
%\begin{equation}
\begin{multline}
    \mathcal H  = \frac U2 \sum_{i,j}\hat n_{i,j}(\hat n_{i,j}-1) \\
            - \sum_{i,j}\left(Ke^{-i \phi_{i,j}}\hat a_{i+1,j}^\dagger \hat a_{i,j} + J \hat a^\dagger_{i,j+1}\hat a_{i,j} + \mathrm{h.c}\right),
\end{multline}
%\end{equation}
where $U$ is an on-site, pairwise repulsive interaction energy and $K$ and $J$ are tunneling amplitudes between nearest neighbors on the lattice. $\hat a^\dagger_{i,j}$, $\hat a_{i,j}$, and $\hat n_{i,j}$ are the creation, annihilation, and number operators for site $(i,j)$, where $i \in \mathbb Z$ and $j \in \{0, 1\}$ for a ladder geometry. The spatially varying complex tunneling phases, $\phi_{i,j}$, are realized through the combination of a magnetic field gradient to energetically detune nearest neighbors and a running lattice to drive Raman transitions to restore tunneling between these sites~\cite{aidelsburger_experimental_2011,miyake_realizing_2013}. For a loop around a unit cell, these tunneling phases result in a non-trivial net phase $\Phi$, yielding an effective magnetic field by playing the role of the Aharonov-Bohm phase acquired by a charged particle in a real magnetic field. The flux $\Phi$ is controlled in our system by the angle between the running lattice and the static lattice on which the atoms reside. Because the Raman lattice is projected through the objective, as shown in Figure~\ref{fig:experiment_schematic}a, we are able to dynamically tune the effective magnetic field from the weak to strong field limits within a single experiment, without having to change the laser wavelengths (see Methods).

Finally, we utilize a digital micromirror device in a Fourier plane of our projection path to superimpose arbitrary optical potentials~\cite{zupancic_ultra-precise_2016}, which we use in state preparation and to confine evolution to a ${2 \times N }$ ladder region. For the experiments that follow, we operate in the regime where $K/h = 11(1) \, \text{Hz}$, $J/h = 34.1(6) \, \text{Hz}$, and $U/h = 131.2(6) \, \text{Hz}$, with $h$ being Planck's constant. The tunneling rates and interaction energy are calibrated using modulation spectroscopy (see Methods). The spatially varying phase can be expressed $\phi_{i,j} = \pi \cdot i + \Phi \cdot j$. 

%\section{Single particle chiral dynamics}
\begin{figure}[!htb]
    \centering
    \includegraphics[width=\columnwidth]{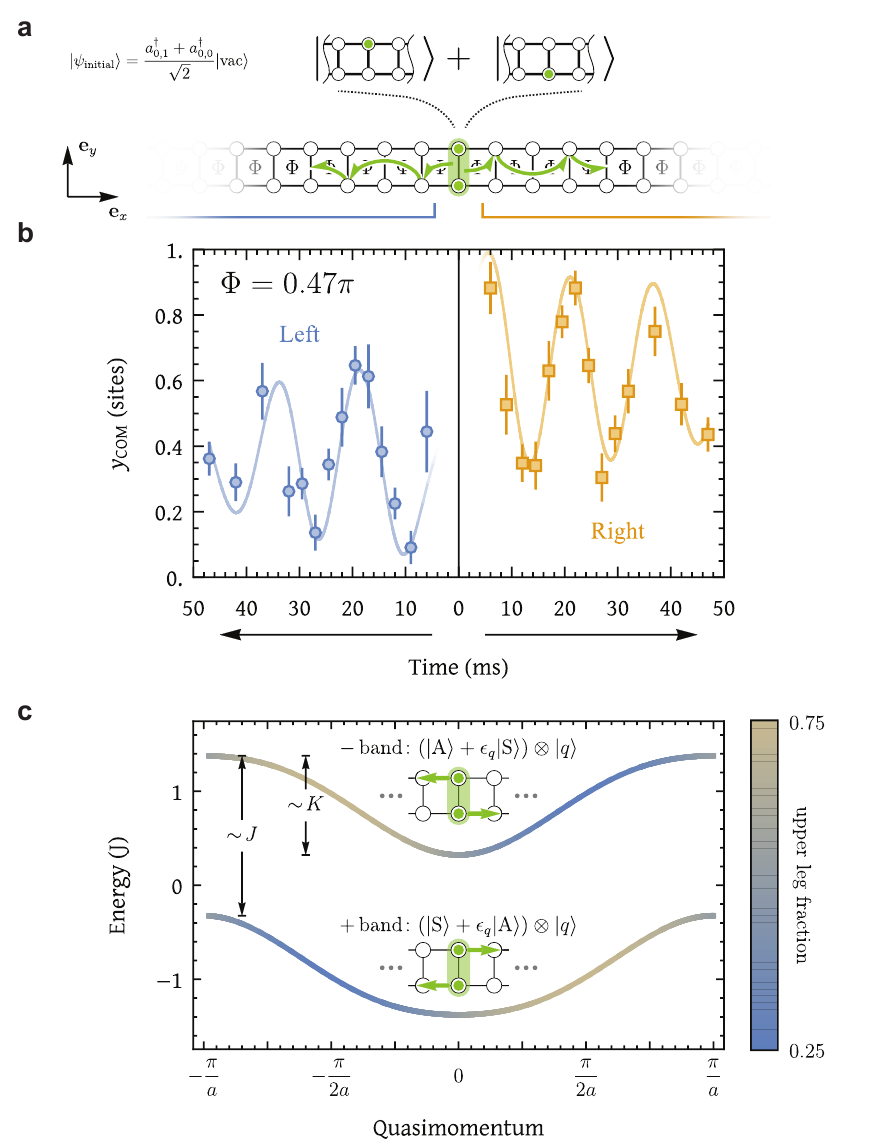}
    \caption{\label{fig:single_particle} 
    \textbf{a},	In the presence of a gauge field, a particle delocalized across a single rung exhibits chiral motion (an exemplary trajectory is indicated by the green arrows). 
	\textbf{b}, To quantify the chiral dynamics, we track the time evolution of the $x$ and $y$ components of the center-of-mass for the left (blue) and right (orange) halves of the system for a flux of $\Phi = 0.47 \pi$. Population on the initial rung is always excluded from the center-of-mass determination. While the particle symmetrically delocalizes over multiple sites in the $x$ direction (see Methods), its motion in the $y$ direction is asymmetric with respect to the central rung. The data is matched well by a simulation using exact diagonalization of the Harper-Hofstadter model, performed for our system with the tunnelings fit to $K/h = 12(1) \, \text{Hz}$ and $J/h = 29(2) \, \text{Hz}$. Due to technical errors in the initial state preparation (see Methods), the chiral trajectory is slightly modified from the case of a symmetric superposition.
    \textbf{c},	Band structure computed for the simulation parameters used in \textbf{b}. The spectrum exhibits two bands, with the upper leg population fraction of each eigenstate encoded by color. In each band, there is a correlation between the sign of the group velocity and the leg towards which the particle is biased. However, the sign of the correlation differs between the two bands, giving rise to chiral bands of opposite sign, which we denote by $+$ and $-$. The chirality of each band is shown as an inset, where particles preferentially stick to one of the legs depending on the propagation direction. The initial state prepared in \textbf{b} preferentially populates the lower band, resulting in chiral dynamics associated with that band.}
\end{figure}

We first probe the chiral band structure of this system by studying single-particle dynamics~\cite{huegel_chiral_2014}. In these experiments, an atom is repeatedly prepared in a known initial state and the average density distribution is obtained for several evolution times. We start from a Mott insulator and remove all but a single atom~\cite{preiss_strongly_2015,islam_measuring_2015}. By means of a Landau-Zener sweep, this atom is delocalized to create the ground state of the central rung subsystem, ${|\psi_\text{initial}\rangle_{1p} = \frac{1}{\sqrt{2}}(a_{0, 1}^\dagger + a_{0, 0}^\dagger) \vert \text{vac} \rangle \equiv \hat{a}^\dagger_{0, \text S} \vert \text{vac} \rangle}$. A sweep in the opposite direction would prepare ${\hat{a}^\dagger_{0,\text A} \vert \text{vac} \rangle = \frac{1}{\sqrt{2}}(a_{0, 1}^\dagger - a_{0, 0}^\dagger) \vert \text{vac} \rangle}$. From the initial state $\hat{a}^\dagger_{0, \text S} \vert \text{vac} \rangle$, we investigate the effect of the artificial magnetic field on the atom's propagation dynamics after suddenly reducing the lattice depth in the $x$ direction~\cite{stuhl_visualizing_2015,mancini_observation_2015,an_direct_2016}. 

At zero flux, the particle's motion is separable and, hence, we would expect no dynamics along the $y$ direction since we prepared the atom in the ground state along this dimension. However, a particle's coupling to a magnetic field gives rise to a non-separable Hamiltonian, yielding chirality and multi-dimensional dynamics in the atom's motion. In Figure~\ref{fig:single_particle}b, the $y$ components of the center-of-mass, $y_\text{COM}$, for the left and right halves of the systems are plotted as a function of evolution time. The density in both parts of the system oscillates between the upper and lower leg of the ladder, but these oscillations occur out of phase with each other and are biased towards opposite legs. The population that propagates to the right (left) is initially biased towards the upper (lower) leg of the ladder --- a behavior that is reminiscent of skipping orbits. 
 
We explain the chiral trajectories observed in the quantum walk using the ladder's band structure. An isolated rung subsystem admits eigenstates with antisymmetric and symmetric superpositions of the atom occupying each constituent site. In the limit that $J > K$, these eigenstates hybridize with plane wave states running along the legs of the ladder such that there are two sub-bands with non-zero width ${\sim} K$, split by an energy ${\sim} J$ (Figure~\ref{fig:single_particle}c). Each band is composed of Bloch states of quasimomentum, $q$, where the rung subsystems define the unit cells of a one-dimensional lattice. The population of each site in the rung subsystems, color coded in Figure~\ref{fig:single_particle}c, depends on the quasimomentum of the eigenstate.

We identify the bands as ${+}$ or ${-}$ due to their symmetric and antisymmetric character in the unperturbed ${K=0}$ limit~\cite{huegel_chiral_2014} . In the ${+}$ band, a population imbalance towards the upper (lower) leg of the ladder is associated with a rightward (leftward) group velocity, giving rise to chiral behavior. An analogous analysis confirms that the ${-}$ band exhibits chirality of the opposite sign. The initial state shown in Figure~\ref{fig:single_particle}a more heavily populates the ${+}$ band, resulting in the chiral behavior associated with that band. The converse would occur for the initial state $\hat{a}^\dagger_{0,\text A} \vert \text{vac} \rangle $, resulting in chiral dynamics of opposite sign. 

\begin{figure}[!htb]
    \centering
    \includegraphics[width=\columnwidth]{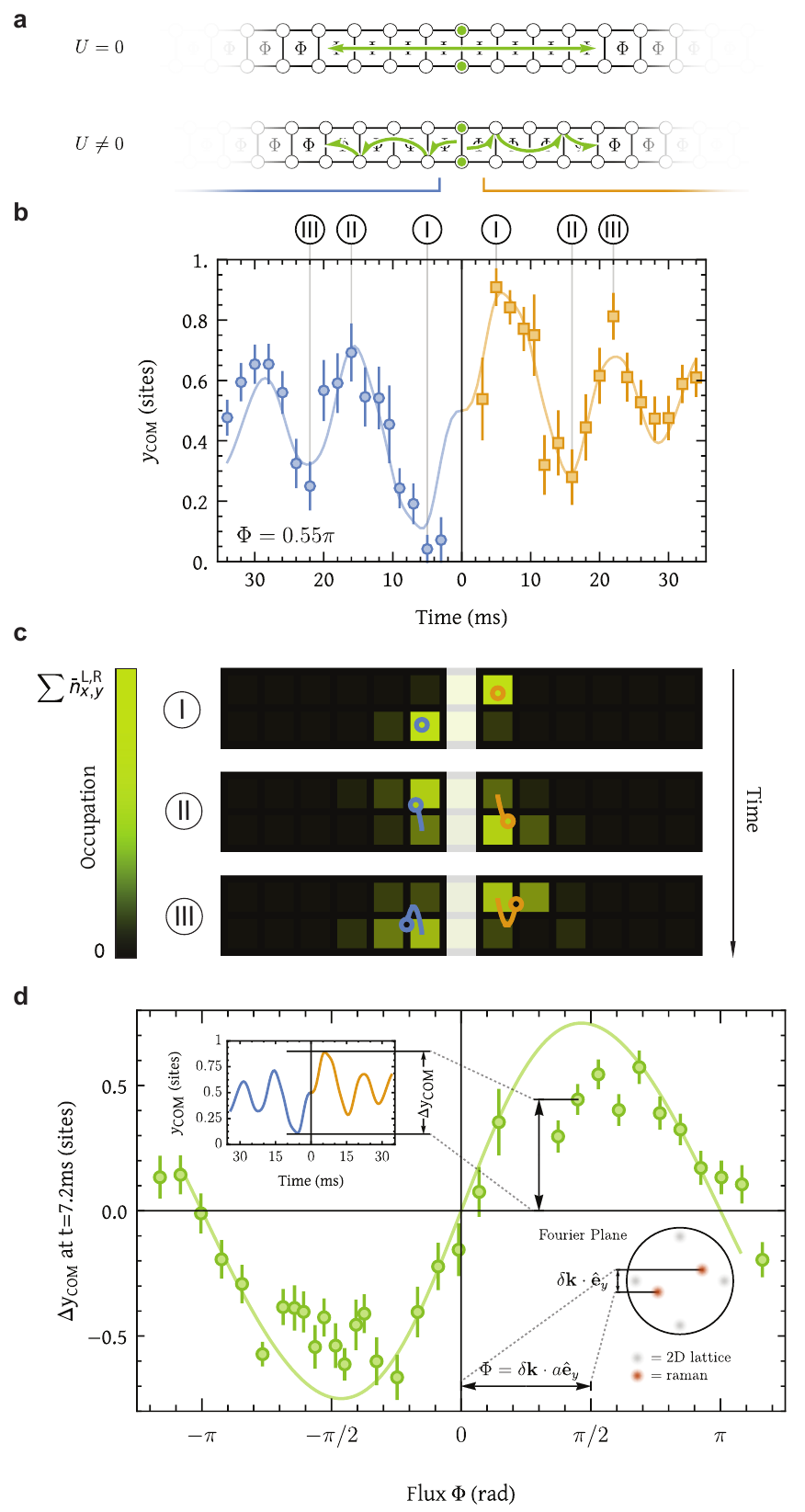}
    \caption{\label{fig:two_particle}
    \textbf{a}, We initialize a pair of particles onto opposite sites of the central rung and track the density distribution in the presence of a gauge field. The interactions give rise to chirality in the propagation dynamics. 
    \textbf{b}, Chirality manifests in the difference in $y_\text{COM}$ for the left (blue) and right (orange) halves of the system. The solid line results from an exact diagonalization at $\Phi = 0.55 \pi$ and $\{U, J, K\}/h = \{131.2(6), 34.1(6), 11(1)\} \, \text{Hz}$.
    \textbf{c}, Exemplary density distributions for times indicated in \textbf{a}. The saturation of a square indicates the occupation probability of that lattice site normalized to the population in its half of the system, left or right. Blue and orange circles indicate the position of the center-of-mass for the left and right halves, respectively. Solid lines trace the evolution of the center-of-mass up to the point of measurement.
    \textbf{d}, We quantify the amount of chirality by the shearing at $t=7.2 \, \text{ms}$, corresponding roughly to the first maxima in the center-of-mass evolution (inset).}
%Our data shows that chirality is absent for the case of zero flux when the gauge field is switched off, as well as at $\pi$ flux, highlighting the role of the gauge field in the chiral dynamics.}
\end{figure}

%\section{Interaction-induced chirality}
We study the interplay between interactions and the synthetic gauge field by preparing two bosons on the two neighboring sites of the central rung in the ladder. We can express this initial state $\vert \psi_\text{initial} \rangle_{2p}$ in terms of the single-particle states discussed in the prior section as $\vert \psi_\text{initial} \rangle_{2p} = a^\dagger_{0,1} a^\dagger_{0,0} |\text{vac}\rangle = \frac1{\sqrt{2}}((a_{0,\text S}^\dagger)^2-(a_{0,\text A}^\dagger)^2) |\text{vac} \rangle$. This decomposition shows that the upper and lower chiral bands of Figure~\ref{fig:single_particle}c are equally populated. Hence, despite the presence of a gauge field, such a state would exhibit no chirality in the non-interacting limit because there is equal weight in bands of opposite chirality and these weights are preserved in time since the non-interacting eigenstates are products of the single-particle eigenstates. In contrast, we observe clear chiral orbits as the particles evolve from $\vert \psi_\text{initial} \rangle_{2p}$ in our experiment where interactions play a critical role, as shown in Figure~\ref{fig:two_particle}. By varying the flux with our projective scheme, we see that the chirality observed is present in the two-particle trajectories whenever the applied flux induces chirality in the single-particle bands (Figure~\ref{fig:two_particle}c), specifically, when the flux $\Phi$ is neither zero nor $\pi$. Our data thus establishes that the observed chiral dynamics depend on both the interactions and the applied synthetic gauge field.

\begin{figure}[!htb]
    \centering
    \includegraphics[width=\columnwidth]{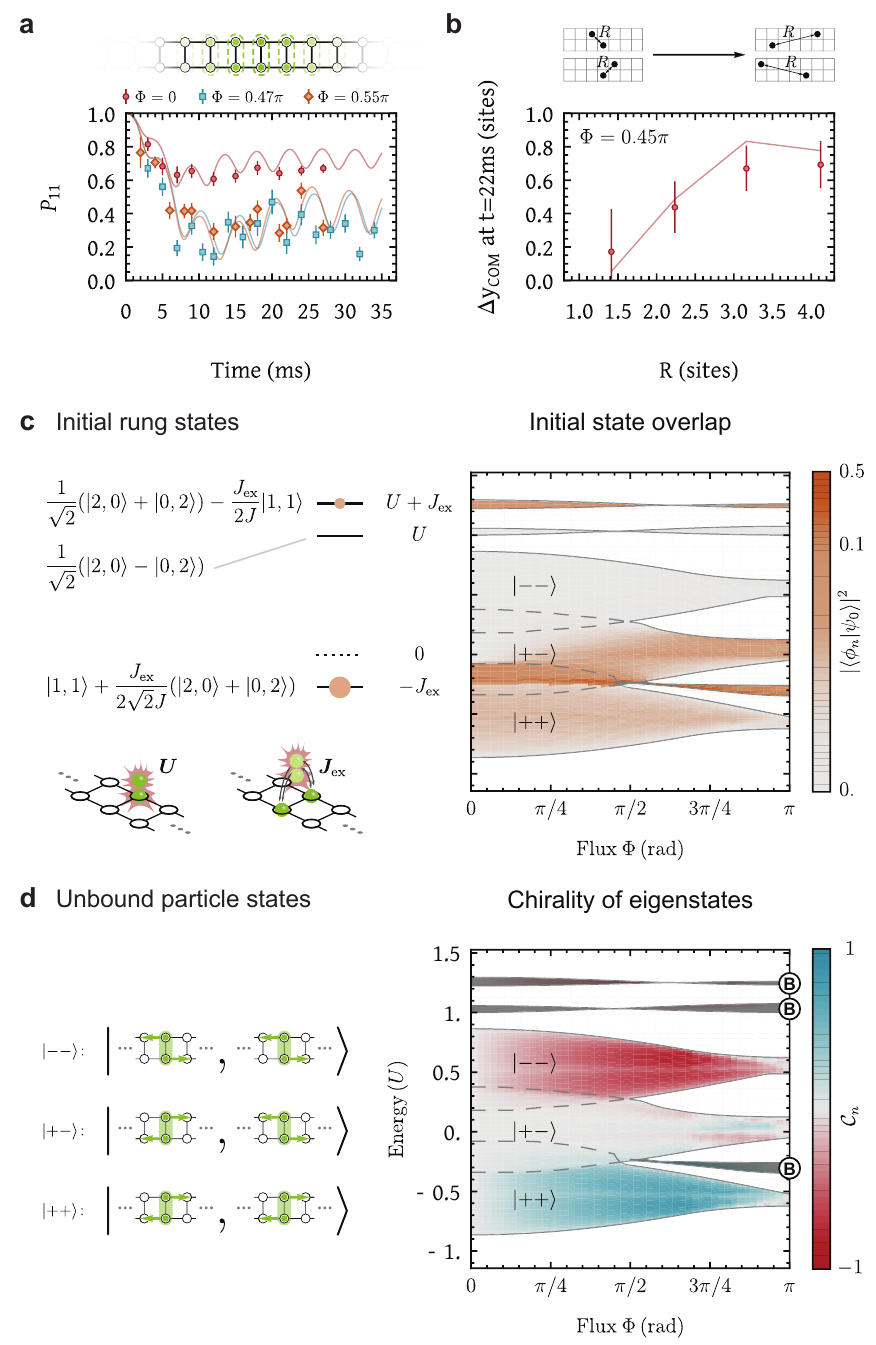}
    \caption{\label{fig:bound_states} The interacting two-particle dynamics give rise to both bound states and scattering states, the interplay of which leads to chirality. 
     \textbf{a}, We plot the probability $P_{11}$ that the particles are on different sites of the same rung as a function of time and flux; this quantity provides a measure of the population in the lowest energy bound state (see panel \textbf{d}). The dynamics in $P_{11}$ are shown for fluxes of $\Phi=0$, $\Phi=0.47 \pi$, and $\Phi=0.55 \pi$. 
     \textbf{b}, The degree of shearing, $\Delta y_\text{COM}$, following $22 \, \text{ms}$ of evolution in the ladder is plotted as a function of the inter-particle spacing, i.e. the average $\Delta y_\text{COM}$ when the atoms are a distance $R$ apart. The shearing, a proxy for the chirality, suggests that the scattering states where particles are further separated contribute most to the chirality. 
     \textbf{c}, We show (left) the population of the eigenstates of a given rung in the $K=0$ limit and (right) the full many-body spectrum as a function of flux $\Phi$, given the calibrated $J$, $K$, and $U$. In the latter, the color encodes the overlap between our initial state and these many-body eigenstates as a function of $\Phi$, where the boundaries of the different eigenstate regions of interest are differentiated by the gray lines. The regions enclosing $|{-}{-}\rangle$, $|{+}{-}\rangle$, and $|{+}{+}\rangle$ are then largely described by product states of the unbound eigenstates.
    \textbf{d}, Chiral character of the many-body eigenstates. The $+$ and $-$ labels describe what single-particle bands mostly describe the colored bands in this plot. The color denotes the chirality, $\mathcal C_\mathrm{n}$, of the eigenstates (see Methods) within the bands and the dark gray regions refer to the eigenstates that compose the bound states in the ladder system. \vspace{-0.5cm}}
\end{figure}

To derive a physical picture for how the interactions introduce chirality into the dynamics of the two-particle quantum system, we delineate two classes of eigenstates. There is one class where the bosons are bound through interactions and another class of unbound scattering states where the particles are largely independent and approximately equal to product states of the single-particle eigenstates. This latter category we refer to as $\ket{{+}{+}}$, $\ket{{+}{-}}$, or $\ket{{-}{-}}$ states, depending on the bands populated by the bosons. Since the bosons in our initial state are close to each other, we expect that there can be sizable overlaps with both types of eigenstates for our experimental parameters. To study the decomposition of the initial state experimentally, we measure the probability $P_{11}$ for two bosons to occupy neighboring sites of the same rung anywhere in the system as both a function of time and gauge field strength (Figure~\ref{fig:bound_states}a). Indeed, even at long times we find a sizable, flux-dependent probability for the bosons to remain close to each other, indicating the presence of a bound state.

To analyze which trajectories contribute most to the chiral signal, we study the shearing amplitude (as in Figure~\ref{fig:two_particle}) as a function of the inter-particle distance. At long times, we expect the population in eigenstates of largely free-particle (bound) character to yield atoms that are farther apart (closer together). We find that the shearing increases for bosons farther apart (Figure~\ref{fig:bound_states}b), suggesting that the populated, unbound scattering states contribute mostly to the observed chirality. This implies that there is an imbalance in the populations of $\ket{{+}{+}}$ and $\ket{{-}{-}}$ states, which are of opposite chirality. Given the stationary and equal population of the chiral bands in the non-interacting case, the imbalance in these observed dynamics must be induced by the interactions. We confirm this conclusion by calculating numerically the overlaps of our initial state with the eigenstates of the full interacting Hamiltonian and the associated chirality, $\mathcal{C}_\mathrm{n}$ (see Methods), of these eigenstates (Figure~\ref{fig:bound_states}c,d). In addition, a short-time perturbative expansion shows that $y_{\mathrm{COM}}(t) = \pm a(t/\hbar)^{5} (KJ)^2 U \sin(\Phi)$ to leading order in the evolution time $t$ (see Methods) for the right (+) and left (-) side of the ladder, indicating that the dynamics depend both on $\Phi$ and the interactions $U$.

%Although the observed chirality arises from population in non-interacting states, the population imbalance between states with two bosons in the lower ($\ket{{+}{+}}$) and upper ($\ket{{-}{-}}$) chiral bands requires interactions. 

To illustrate the process by which interactions induce a population imbalance between ($\ket{{+}{+}}$) and upper ($\ket{{-}{-}}$) scattering states, we analyze the eigenstates of two interacting bosons on the central rung and study how they hybridize with delocalized states as the tunnel coupling $K$ is introduced. In the limit $U \gg J$, two of the three two-boson eigenstates involve double occupancy of one site, causing an energy shift of order $U$ (Figure~\ref{fig:bound_states}c). These states are largely detuned from the delocalized states and do not influence the dynamics. There is a third eigenstate near zero energy, with large overlap to our initial state. Due to the finite $J$ and $U$, this eigenstate is shifted down in energy by ${J_\text{ex} = 4J^2/U}$ through a super-exchange process (Figure~\ref{fig:bound_states}c). For $U=0$, the tunnel coupling $K$ along the legs hybridizes this rung eigenstate equally with states $\ket{{-}{-}}$ and $\ket{{+}{+}}$. However, because of the downward energy shift of $J_\text{ex}$ for $U\neq0$, the state is energetically closer to $\ket{{+}{+}}$ states and hybridizes primarily with this chiral band (Figure~\ref{fig:bound_states}c). The interactions also create a delocalized bound state when $J_\text{ex} \sim K$, corresponding to a particle on each site of a rung somewhere in the system, which is reflected by the non-vanishing $P_{11}$. Lastly, as the magnetic flux $\Phi$ is increased, the motion of the particles in the $x$ and $y$ directions becomes increasingly coupled by the gauge field, leading to additional hybridization with the $\ket{{+}{-}}$ states (Figure~\ref{fig:bound_states}c). This effect is also in agreement with the observed reduction of $P_{11}$ at long times when introducing the gauge field (Figure~\ref{fig:bound_states}a), because we expect an increase in the size of the bound state caused by the coupled motion in the $x$ and $y$ direction. %This last effect removes population from the bound sector, yielding the observed reduction of $P_{11}$ at long times when introducing the gauge field compared to when $\Phi=0$  (Figure~\ref{fig:bound_states}a). 

In conclusion, the combined effect of interactions and a synthetic gauge field leads to our observation of chirality in the multi-particle dynamics of bosons in a $2 \times N$ ladder. Our observations depend on being in a regime where the interactions are neither vanishing nor infinite --- in either of these limiting cases, there is no exchange-energy shift $J_\text{ex}$, leading to symmetric populations in the two chiral sectors. Our results suggest that the mere presence of interactions may facilitate the preparation of correlated many-body states in a chiral band, which represents a key challenge for realizing exotic fractional quantum Hall states of neutral atoms in a topologically non-trivial band. Our apparatus is capable of performing future experiments in a regime where further exotic physics is expected.

\begin{acknowledgments}
We would like to acknowledge helpful conversations with Monika Aidelsburger, Ignacio Cirac, Eugene Demler, Manuel Endres, Michael Foss-Feig, Nathan Gemelke, Daniel Greif, Wolfgang Ketterle, Ruichao Ma, Hoi Chun Po, Jonathan Simon, and Ashvin Vishwanath. Furthermore, we are supported by grants from the National Science Foundation, Gordon and Betty Moore Foundation's EPiQS Initiative, an Air Force Office of Scientific Research MURI program, an Army Research Office MURI program, and the NSF Graduate Research Fellowship Program (MNR).
\end{acknowledgments}

\bibliography{main_text_arXiv}

%\appendix 
\setcounter{figure}{0}
\makeatletter 
    \renewcommand{\thefigure}{S\@arabic\c@figure} 
\makeatother 
\section*{Methods}
\section{Theory for Quench Dynamics in a Harper Ladder}

\subsection{Bloch Bands and Wannier Functions of the Harper Ladder}

%%%%%%%%%%%%%%%%%%%%%%%%%%%%%%%%%%%%%%%%%%%%%%%%%%%%%
In this section we consider the non-interacting limit where $U=0$. Despite the presence of the magnetic flux $\Phi$ per plaquette, a gauge choice can be made for the ladder geometry with a two-site unit cell corresponding to an individual rung. Now we derive the cell-periodic Bloch wavefunctions $\ket{u_\pm(q)}$ of the resulting two bands. After deriving their dispersion relations $\epsilon_\pm(q)$, we discuss particle-hole symmetry and calculate the corresponding Wannier functions $w_{\pm}(x,y)$. 

%%%%%%%%%%%%%%%%%%%%%%%%%%%%%%%%%%%%%%%%%%%%%%%%%%%%%
\begin{figure*}[!htb]
    \centering
    \includegraphics{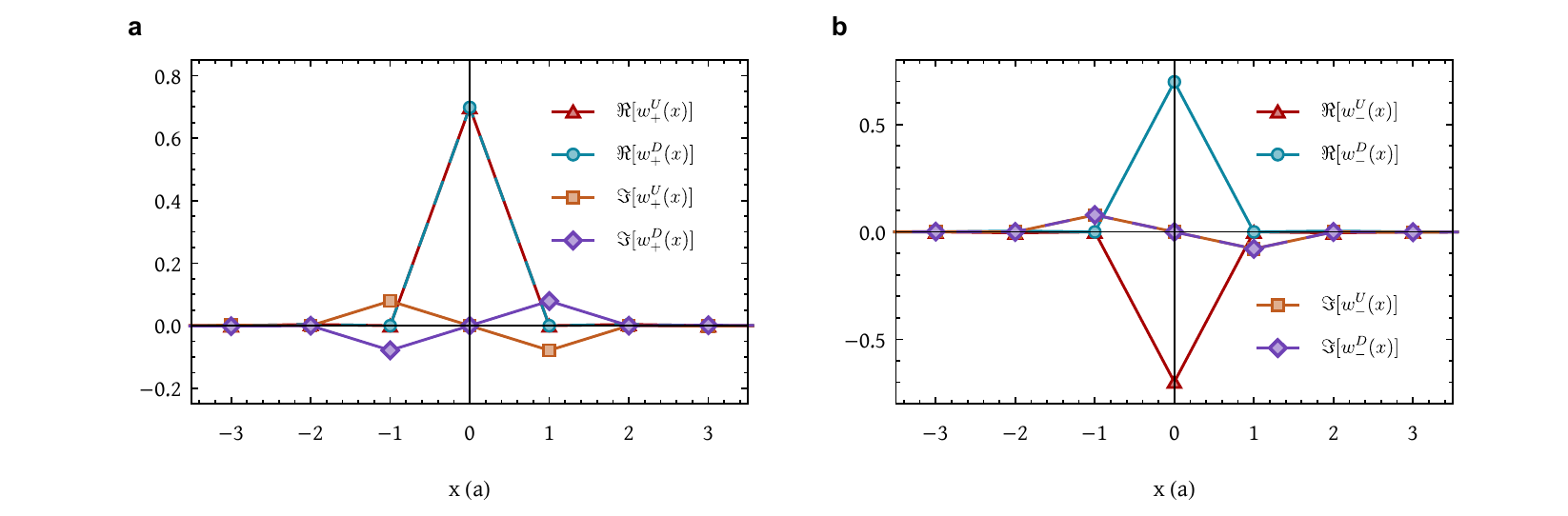}
    \caption{\label{fig:Wannier} \textbf{a}, The real and imaginary parts of the Wannier function $w_+^\mu(x)$, for $\mu=\rm U, D$, is shown, for $J= 34.1 \, \text{Hz}$ $K=11.4 \, \text{Hz}$, and $\Phi=\pi/2$. \textbf{b}, The real and imaginary parts of the Wannier function $w_-^\mu(x)$, for $\mu=\rm U, D$, is shown, for $J=2 K$ and $\Phi=\pi/2$.}
\end{figure*}
%%%%%%%%%%%%%%%%%%%%%%%%%%%%%%%%%%%%%%%%%%%%%%%%%%%%%

%%%%%%%%%%%%%%%%%%%%%%%%%%%
\subsubsection{Bloch Hamiltonian}
%%%%%%%%%%%%%%%%%%%%%%%%%%%
To derive the Bloch Hamiltonian, we make a gauge choice where the tunneling elements describing hoppings in the positive $x$-direction along the lower (upper) leg of the ladder have a complex phase $\Phi/2$ ($- \Phi/2$). The resulting unit-cell is defined by the two sites of a rung of the ladder, with the two states $\ket{ {\rm U} }, \ket{ {\rm D} }$ corresponding to the upper (``up") and lower (``down") leg of the ladder. Notably, the unit-cell is not related to the magnetic unit-cell which encloses an integer number of magnetic flux quanta, reflecting the one-dimensional character of the ladder system.

Without the rung coupling $J$, the Bloch Hamiltonians of the two decoupled legs are given by $- 2 K  \cos (q \pm \Phi/2)$ where the phases $\pm \Phi/2$ only lead to a shift of the two quasimomenta $q$. In terms of Pauli matrices $\hat{\sigma}^{x,z}$, defined in the basis $\ket{ {\rm U} }, \ket{ {\rm D} }$ by the relations $\hat{\sigma}^z \ket{ {\rm U} } = \ket{ {\rm U} }$ and $\hat{\sigma}^z \ket{{\rm D}} = - \ket{{\rm D}}$, this Bloch Hamiltonian can be written $\H(q) = - K ( \nu_+(q) + \nu_-(q) \hat{\sigma}^z )$, where 
\[\nu_\pm(q) = \cos (q + \Phi/2) \pm \cos (q - \Phi/2).\]
The coupling $J$ between the two legs is described by an additional term $- J \hat{\sigma}^x$. The final Bloch Hamiltonian of the ladder system in the presence of the artificial gauge field, for which $\H(q) \ket{u_\pm(q)} = \epsilon_\pm(q) \ket{u_\pm(q)}$, is thus given by
\[\H(q) = - K \nu_+(q) - \l J \hat{\sigma}^x + K \nu_-(q) \hat{\sigma}^z \r.\]

The cell-periodic Bloch functions and the corresponding energies can easily be calculated,
\begin{flalign}
\ket{u_+(q)} &= \cos \frac{\vartheta_{q}}{2} \ket{{\rm U}} + \sin \frac{\vartheta_{q}}{2} \ket{{\rm D}},\\
\ket{u_-(q)} &= \cos \frac{\vartheta_{q}}{2} \ket{{\rm D}} - \sin \frac{\vartheta_{q}}{2} \ket{{\rm U}},
\end{flalign}
where the angle $\vartheta_{q}$ is defined by
\[J \vec{e}_x + K \nu_-(q) \vec{e}_z  = \sqrt{J^2 + K^2 \nu_-^2(q) } \l \sin \vartheta_{q} \vec{e}_x + \cos \vartheta_{q} \vec{e}_z \r.\]
The dispersion relations of the two bands are given by
\[ \epsilon_\pm(q) = - K \nu_+(q) \mp \sqrt{J^2 + K^2 \nu^2_-(q)}.\]
These dispersion relations are shown in Fig.~3c in the main part of the paper.

Physically the two bands ($\pm$) can be understood as hybrids of two counter-propagating chiral edge states. Switching between the two kinds of hybridization, which are symmetric ($+$) and antisymmetric ($-$) superpositions of the upper and lower legs of the ladder respectively, can be formalized by the particle-hole operator which we discuss next. 

%%%%%%%%%%%%%%%%%%%%%%%%%%%%%%%%%%%%%%%%%%%%%%%%%%%%%
\begin{figure*}[!htb]
    \centering
    \includegraphics{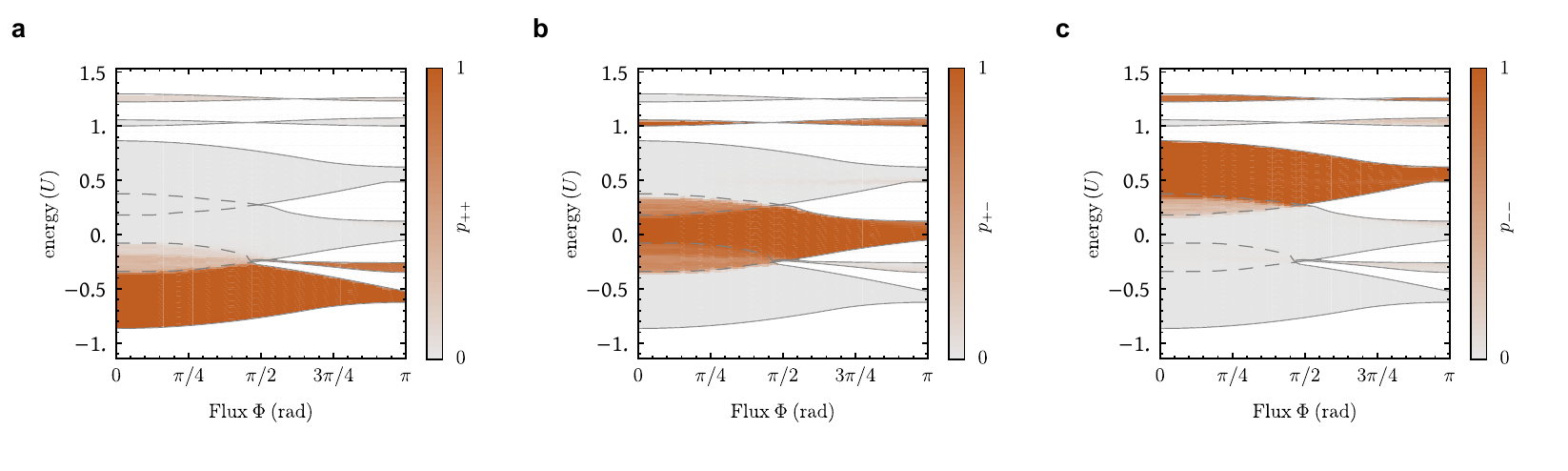}
    \caption{\textbf{Decomposition of interacting into free two-particle eigenstates}: \textbf{a}, The overlap $p_{++}(n)$ is shown (color code) as a function of the eigenenergies $E_n$ and the flux $\Phi$ per plaquette of the synthetic gauge field, see Eq.~\eqref{eq:defpPP}. In \textbf{b} and \textbf{c}, the calculation is repeated for $p_{+-}$ and $p_{--}$ respectively.}
    \label{fig:Decomp2P}
\end{figure*}
%%%%%%%%%%%%%%%%%%%%%%%%%%%%%%%%%%%%%%%%%%%%%%%%%%%%%

%%%%%%%%%%%%%%%%%%%%%%%%%%%
\subsubsection{Particle-hole symmetry}
%%%%%%%%%%%%%%%%%%%%%%%%%%%
The Bloch Hamiltonian $\H(q)$ has particle-hole symmetry. Up to the part of the dispersion relation which is independent of the band index, the Bloch Hamiltonian anti-commutes with $i \hat{\sigma}^y$,
\[\left\{ \H(q) + K \nu_+(q) , i\hat{\sigma}^y \right\} = 0.\]
Therefore by applying $i \hat{\sigma}^y$ the band index of the cell-periodic Bloch function is switched,
\[i \hat{\sigma}^y \ket{u_+(q)} = - \ket{u_-(q)}, \quad i \hat{\sigma}^y \ket{u_-(q)} = \ket{u_+(q)}.\]
In terms of the bosonic operators $\a_{i,j}$ introduced in the main text (where $i$ is the rung index along the ladder and $j=0,1$ corresponds to the two legs on every rung) the particle-hole symmetry is defined by the unitary transformation
\[\hat{C}^\dagger \a_{i,{\rm U}} \hat{C} = - \a_{i,{\rm D}}, \qquad \hat{C}^\dagger \a_{i,{\rm D}} \hat{C} = \a_{i,{\rm U}}.\]
It follows for the operators $\a_{q,\tau}$, which create a boson in the Bloch wave with momentum $q$ along the ladder and in the band indexed by $\tau=\pm$, that 
\[\hat{C}^\dagger \a_{q,\pm} \hat{C} = - \a_{q,\mp},\]
and it holds for $U=0$
\begin{multline*}
    \hat{C}^\dagger \H \hat{C} = \\
    \sum_{q} \sum_{\tau=\pm}   \ad_{q,\tau} \ad_{q,\tau}  \l - K \nu_+(q) +\tau  \sqrt{J^2 + K^2 \nu^2_-(q)}  \r.
\end{multline*}

%%%%%%%%%%%%%%%%%%%%%%%%%%%
\subsubsection{Wannier functions}
%%%%%%%%%%%%%%%%%%%%%%%%%%%
The Wannier functions $w_\tau$ corresponding to the two bands ($\tau = \pm$) are defined by 
\begin{equation}
w_\tau^\mu(x-x_j) = \frac{1}{\sqrt{2 \pi}} \int_{- \pi}^\pi dq ~ e^{i q (x-x_j)} u_\tau^\mu(q),
\label{eq:defWannier}
\end{equation}
where $x,x_i \in \mathbb{Z}$ label one-dimensional lattice sites along the ladder and $\mu=\rm U,D$ denotes the two legs of the ladder. Note that $x$ and $\mu$ represent the coordinates of the Wannier function $w_\tau$ corresponding to the band labeled by $\tau$; similarly $u_\tau^\mu(q)$ labels the component corresponding to state $\ket{\mu}$ of the Bloch wavefunction $\ket{u_\tau(q)}$.

\emph{Numerical results:}
The Fourier transformation Eq.~\eqref{eq:defWannier} can be done numerically, and we find for all values of $\Phi$ that the Wannier functions are very well localized on a single rung of the ladder. For $\Phi \neq 0$ there is a small amplitude on the neighboring rungs, but even for the largest values of $\Phi$ less than $10 \%$ of the orbit leaks onto neighboring rungs.

As an example, we show the Wannier functions $w_+^\mu(x)$ and $w_-^\mu(x)$ for the case $\Phi=\pi/2$ in Figs.~\ref{fig:Wannier}a and \ref{fig:Wannier}b. To a rather good approximation they are given by 
\[\ket{w_\pm(x-x_i)} \approx \delta_{x,x_i} \l \ket{{\rm D}}  \pm \ket{{\rm U}} \r / \sqrt{2},\]
i.e. in terms of bosonic operators:
\begin{equation}
 \ad_{i,\pm} \approx \l \ad_{i,{\rm D}} \pm \ad_{i,{\rm U}} \r / \sqrt{2}.
\label{eq:approxApmj}
\end{equation}
While this expression is exact in the absence of the gauge field, $\Phi=0$, the approximation is not sufficient to understand the chirality in the system, which arises from the coupling of the motion along and transverse to the ladder.

\emph{General properties:}
From Eq.~\eqref{eq:defWannier} a few general properties of the Wannier functions can be derived exactly using the symmetries of the system. The first concerns the relation between $\ket{w_+}$ and $\ket{w_-}$, which follows from the particle-hole symmetry of the system,
\[w_\tau^{\rm U}(x) = \tau ~ w_{- \tau}^{\rm D}(x), \quad \tau = \pm 1.\]
Second we discuss how the values of the Wannier functions on different legs are related. Using that the system is invariant under simultaneous exchange of the upper and lower legs, $\rm U \leftrightarrow D$, and inversion of the flux, $\Phi \to - \Phi$, we can show that
\[w_\tau^{\rm U}(x) = \tau \l w_\tau^{\rm D} \r^* = - (w_{-\tau}^{\rm U}(x) )^*.\]
Finally we can also derive the relations of Wannier functions at $\pm x$. Because the system is invariant under simultaneous spatial inversion and exchange of the upper and lower legs, it follows that
\[w_\tau^{\rm U}(-x) = \tau w_\tau^{\rm D}(x), \qquad w_\tau^{\rm D}(-x) = \tau w_\tau^{\rm U}(x).\]

\subsection{Quench Dynamics of the Interacting System -- Physical Picture}
To understand the emergence of chirality in the quantum walk of two interacting bosons, we calculate numerically the decomposition of the initial state $\ket{\psi_{\rm initial}}$ into the eigenstates $\ket{\phi_n}$ of the system,
\[\ket{\psi_{\rm initial}} = \sum_n c_n \ket{\phi_n}.\]
In Fig. 5 (c) of the main text, we first show the squared amplitudes $|c_n|^2$ plotted against the corresponding eigenenergies $E_n$ as a function of the flux $\Phi$. For better visualization we introduced bins, each of which contains about $10$ neighboring eigenenergies, and summed up all the overlaps $|c_n|^2$ corresponding to each bin. Because the squared amplitudes $|c_n|^2$ are conserved in the dynamics, this decomposition allows to understand the behavior of the system at long times.

%%%%%%%%%%%%%%%%%%%%%%%%%%%%%%%%%%%%%%%%%%%%%%%%%%%%%
\begin{figure*}[!hbt]
    \centering
    \includegraphics{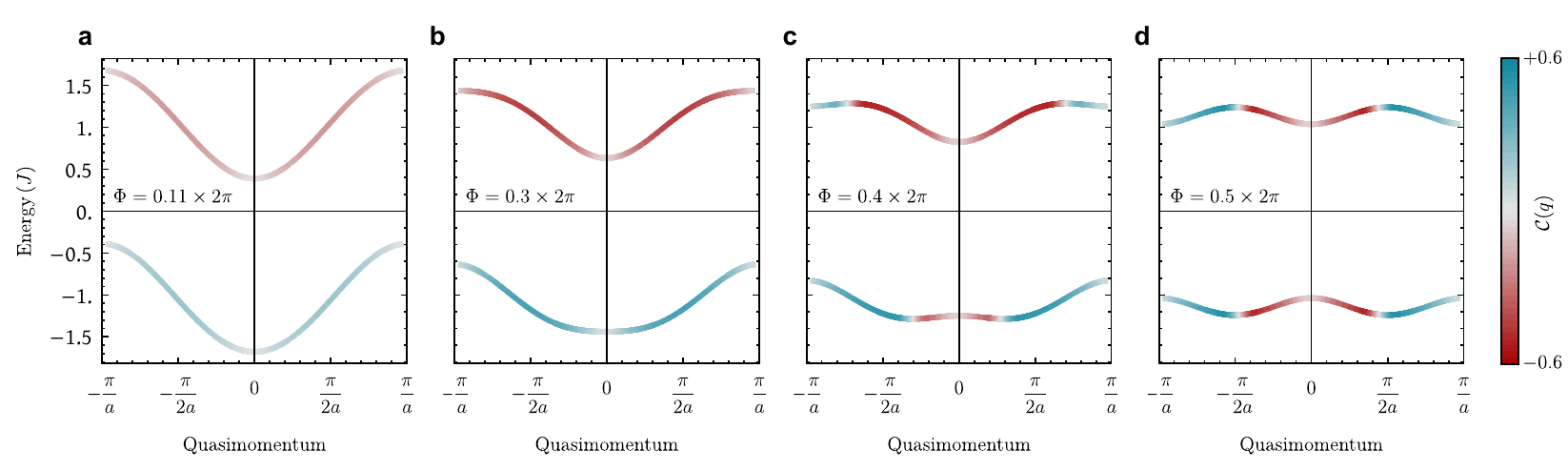}
    \caption{\label{fig:DecompC} \textbf{Chirality of Bloch bands}: The dispersion relations $\epsilon_\pm(q)$ of the two Bloch bands (lower, $+$, and upper, $-$) are plotted, for $J = 2\pi \times 34.1 \, \text{Hz}$ and $K = 2\pi \times 11.4 \, \text{Hz}$ close to the experimental values. The color code indicates the chiralities $\mathcal{C}^\pm(q)$ of the two bands, determined by Eq. \eqref{eq:chiraldef}. Different values of $\Phi$ are considered: \textbf{a}, $0.11 \times 2\pi$, \textbf{b}, $0.3 \times 2\pi$, \textbf{c}, $0.4 \times 2\pi$, and \textbf{d}, $0.5 \times 2\pi$.}
\end{figure*}
%%%%%%%%%%%%%%%%%%%%%%%%%%%%%%%%%%%%%%%%%%%%%%%%%%%%%

To understand the physics of the eigenstates $\ket{\phi_n}$ themselves, we proceed by calculating their decomposition into free two-particle eigenstates,
\begin{multline}
    \ket{\phi_n} =\sum_{k_1,k_2} \biggl[ \frac1{\sqrt2} \phi_{n}^{++}(k_1,k_2)  \ad_{k_1,+} \ad_{k_2,+} \\
    + \frac1{\sqrt2} \phi_{n}^{--}(k_1,k_2) \ad_{k_1,-} \ad_{k_2,-} \\
    + \phi_{n}^{+-}(k_1,k_2) \ad_{k_1,+} \ad_{k_2,-} \biggr] \ket{{\rm vac}},
\end{multline}
where the two-boson wavefunction $\phi_{n}^{++}(k_1,k_2) = \phi_{n}^{++}(k_2,k_1)$ is symmetric (and analogously for $\phi_n^{--}$). Here the operator $\ad_{k,\tau}$ creates a boson at quasimomentum $k$ in the band labeled by $\tau=\pm$. 

To study the influence of interactions on the free eigenstates $\ad_{k_1,\tau_1} \ad_{k_2,\tau_2} \ket{0}$, we calculate their overlaps with the eigenstates $\ket{\phi_n}$ of the interacting system. In Fig.~\ref{fig:Decomp2P} (a) the amplitudes summed over all quasimomenta $k_1$, $k_2$ are shown for $\tau_1=\tau_2=+$,
\begin{equation}
p_{++}(n) = \sum_{k_1,k_2} |\phi_{n}^{++}(k_1,k_2)|^2,
\label{eq:defpPP}
\end{equation}
again as a function of the flux per plaquette $\Phi$ of the synthetic gauge field and the eigenenergy. We used the same binning method for the energies as described for the overlaps $|c_n|^2$ above. In Fig.~\ref{fig:Decomp2P} (b), (c) we repeat this calculation for $p_{--}(n)$ and $p_{+-}(n)$ defined using $|\phi_{n}^{--}|^2$ and $|\phi_{n}^{+-}|^2$, respectively. 

From these figures we draw the conclusion that the energy spectrum of the interacting Hamiltonian consists of two main features. Firstly, we recognize three broad bands of eigenstates (labeled $++$, $+-$, and $--$) which are very similar to the states of free bosons at $U=0$. Most of these states are scattering states, i.e. the two bosons are likely to be far away from each other and their wavefunction can be well approximated by two independent plain waves in that case, $\ad_{k_1,\tau_1} \ad_{k_2,\tau_2} \ket{0}$. This justifies labeling these states by the two quantum numbers $\ket{\tau_1, \tau_2}$ as done in the main text of the paper. In addition to the scattering states, we find three narrow bands of bound states in Fig.~\ref{fig:Decomp2P} (a) - (c) corresponding to repulsively bound pairs with a large effective mass. The energetically lowest branch mostly consists of particles in the lowest ($+$) band. The two upper branches, with energy $\sim U$, mostly correspond to one boson in each band and two bosons in the $-$ band, respectively.

Finally, we describe how we analyzed the chirality of the interacting eigenstates $\ket{\phi_n}$, shown in Fig. 5 (c) in the main text. To this end, we first calculated the chirality of the single-particle eigenstates. For a given quasimomentum $k$ the chirality of band $\tau$ can be defined by
\begin{equation}
\mathcal{C}^\tau(k) = {\rm sign}( v_{\rm g}^\tau(k)) \times \l p^\tau_{\rm U}(k) - p^\tau_{\rm D}(k) \r, 
\label{eq:chiraldef}
\end{equation}
where $ {\rm sign} (v_{\rm g}^\tau(k))$ denotes the sign of the group velocity $v_{\rm g}^\tau(k) = \partial_k \omega_k^\tau$, and $p_{\rm U,D}^\tau(k) = |u_\tau^{\rm U, D}(k)|^2$ are the probability amplitudes for a particle to reside on the upper (U) or lower (D) leg of the ladder. As an example, this chirality is shown color-coded in Fig.~\ref{fig:DecompC} for the experimentally relevant bandstructures. 

To obtain the chirality of the interacting eigenstates $\ket{\phi_n}$, we combine the projection to free two-particle states with the single-particle chirality. This allows us to define the chirality of $\ket{\phi_n}$ as
\begin{multline}
\mathcal{C}_n = \sum_{k_1,k_2} \biggl[  |\phi_{n}^{++}(k_1,k_2)|^2 \l \mathcal{C}^+(k_1) + \mathcal{C}^+(k_2) \r \\
    + |\phi_{n}^{--}(k_1,k_2)|^2 \l \mathcal{C}^-(k_1) + \mathcal{C}^-(k_2) \r \\
    + |\phi_{n}^{+-}(k_1,k_2)|^2 \l \mathcal{C}^+(k_1) + \mathcal{C}^-(k_2) \r \biggr].
\end{multline}
In Fig. 5(d) of the main text we show $\mathcal{C}_n$ as a function of the eigenenergies and the synthetic magnetic field, using the binning method described above.

\subsection{Short-Time Expansion}
\begin{figure*}[!htb]
    \centering
    \includegraphics{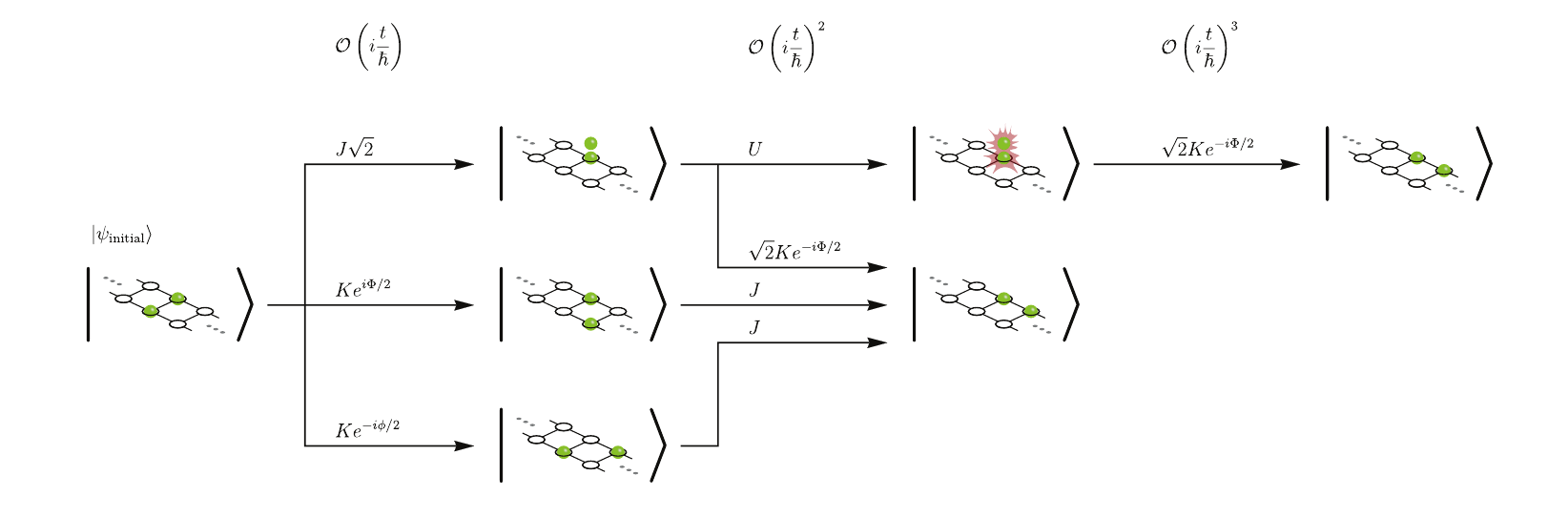}
    \caption{\label{fig:perturbation_figure} Illustration of the leading-order terms contributing to the short-time perturbative expansion: The $\mathcal{O}\left(i\frac{t}{\hbar}\right)^{2}$ paths interfere with the $\mathcal{O}\left(i\frac{t}{\hbar}\right)^3$ path including on-site interactions on the upper leg of the ladder. An equivalent set of interfering paths, with the upper and lower legs exchanged, leads to the leading-order non-vanishing chiral signal scaling like $U t^5 \sin \Phi$, as described in the text.}
\end{figure*}
\begin{figure}[!htb]
    \centering
    \includegraphics{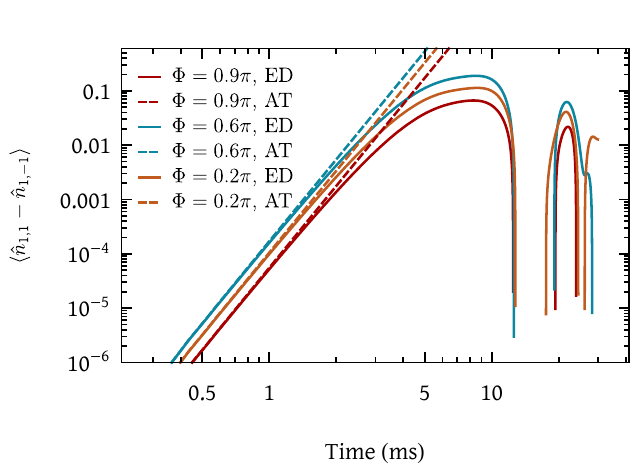}
\caption{\label{fig:logplot} Log plot of both the analytical theory (AT) result $\left\langle\hat{n}_{1,1}-\hat{n}_{1,0}\right\rangle =t^{5}\left(\frac{U}\hbar\right)\left(\frac{KJ}{\hbar^2}\right)^{2}\sin(\Phi)$, which can be obtained using a short-time expansion, and the numerical results obtained from exact diagonalization (ED) for different flux values $\Phi$. Note the strong agreement for $t \lesssim 1 \, \text{ms}$ at all flux and the clear $t^5$ dependence.}
\end{figure}

As mentioned in the text, we compute $\left\langle \hat{n}_{i,1}-\hat{n}_{i,0}\right\rangle$ for the initial state $\ket{\psi(t)} = a_{0,1}^{\dagger}a_{0,0}^{\dagger}\ket{\text{vac}}$, where $i = 0$ is the center of the ladder. Using the Hamiltonian in Eq. (1) and $\ket{\psi(t)} = e^{-i\mathcal{H}t/\hbar}a_{0,1}^{\dagger}a_{0,0}^{\dagger}\ket{\text{vac}}$,  we Taylor expand in $t$ (valid for $t\ll\frac{\hbar}{\max(J,K,U)}$) and compute $\left\langle \hat{n}_{i,1}-\hat{n}_{i,0}\right\rangle$. As expected, when $i=0$, $\left\langle \hat{n}_{0,1}-\hat{n}_{0,0}\right\rangle=0$. Then, for $i = 1$, the relevant non-vanishing terms in $\ket{\psi(t)}$ are
\begin{widetext}
\begin{multline}
    \label{eq:vanishing}
    \ket{\psi(t)} = \ldots -\frac{1}{2!}\left(\frac{t}\hbar\right)^{2} \times 2JK \left(\left(e^{-i\Phi/2}+\cos\frac\Phi2\right)a_{1,0}^{\dagger}a_{0,0}^\dagger + \left(e^{i\Phi/2}+\cos\frac\Phi2\right) a_{1,1}^\dagger a_{0,1}^\dagger\right)\ket{\text{vac}} \\
    + \frac{i}{3!}\left(\frac{t}\hbar\right)^{3} \times 2JKU \left(e^{i\Phi/2}a_{1,1}^\dagger a_{0,1}^\dagger + e^{-i\Phi/2}a_{1,0}^{\dagger}a_{0,0}^{\dagger}\right)\ket{\text{vac}},
\end{multline} 
as illustrated in Figure~\ref{fig:perturbation_figure}. 
This yields ($i=-1$ is similar)
%\begin{align}
%    \left\langle \hat{n}_{1,1}-\hat{n}_{1,0}\right\rangle & = -\left\langle \hat{n}_{-1,1}-\hat{n}_{-1,0}\right\rangle \nonumber \\
%    \label{eq:perturbresult}
%    & = t^{5}\left(\frac{U}\hbar\right)\left(\frac{KJ}{\hbar^{2}}\right)^{2}\sin(\Phi).
%\end{align}
\begin{equation}
    \label{eq:perturbresult}
    \left\langle \hat{n}_{1,1}-\hat{n}_{1,0}\right\rangle  = -\left\langle \hat{n}_{-1,1}-\hat{n}_{-1,0}\right\rangle = t^{5}\left(\frac{U}\hbar\right)\left(\frac{KJ}{\hbar^{2}}\right)^{2}\sin(\Phi).
\end{equation}
\end{widetext}
The odd dependence on the flux, $\Phi$, arising from the on-site interaction occurring on the top versus the bottom leg, indicates that the effect is chiral. For small t, Figure~\ref{fig:logplot} shows good agreement between equation~\ref{eq:perturbresult} and numerical results, convincingly showing a $t^{5}$ dependence.

In the $U\to\infty$ limit, we recompute the short-time expansion with hard-core bosons, and notice the non-vanishing terms in equation \ref{eq:vanishing} require the asymmetry of the particles interacting on the top versus the bottom legs of the ladder. Such a case is not possible in the hard-core boson limit, and up to order $t^{5}$, all terms vanish.

%%%%%%%%%%%%%%%%%%%%%%%%%%%%%%%%%%%%%%%%%%%%%%%%%%%%%

\clearpage

\section{Calibrations}

\subsection{Tunneling Calibration}

\begin{figure}[!htbp]
    \centering
    \includegraphics{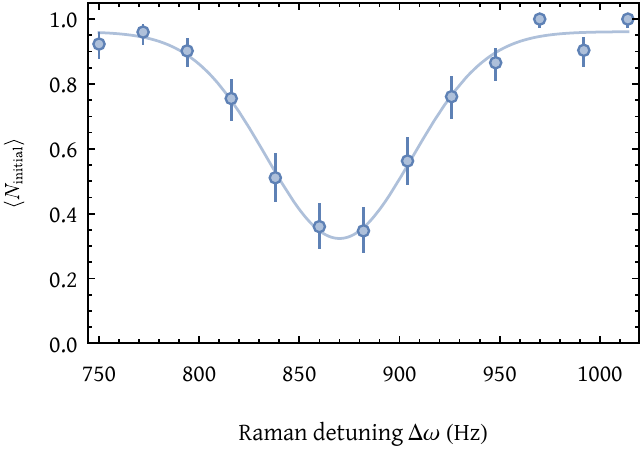}
    \caption{\textbf{Resonant frequency calibration}: Occupation of the initial site in a tilted lattice after $t = 8.5\,\text{ms}$ of evolution plotted against the detuning of the Raman beams.}
    \label{fig:ResonancesCal}
\end{figure}

% Resonances calibration
We calibrate the tunneling rates in the presence of the Raman lattice and the overall lattice tilt.
The tilt is engineered by applying a physical magnetic field gradient along the leg dimension of the ladder.
We first find the frequency of the Raman lattice that restores tunneling along this leg dimension, which is also referred to as the $x$ dimension throughout this text.
As discussed in the main text, the Raman lattice is created by projecting a pair of beams with relative detuning $\Delta \omega = E$ through our microscope objective, where $E$ is the energy offset between neighboring lattice sites that arises from the applied lattice tilt.
Experimentally, we find $\Delta \omega$ by initializing a single atom in a one-dimensional lattice and observing the occupation of the original site as a function of detuning of the two beams.
The obtained same-site occupation after a short time $t = 8.5\,\text{ms} \ll h /K$ is shown in Fig.~\ref{fig:ResonancesCal}.
The detuning where the site occupation is minimal corresponds to the point where resonant tunneling is restored.
For all experiments we set $\Delta \omega / 2 \pi \approx 870\,\text{Hz}$, which is the minimum of a Gaussian fit to the data.

\begin{figure*}[!htbp]
    \centering
    \includegraphics{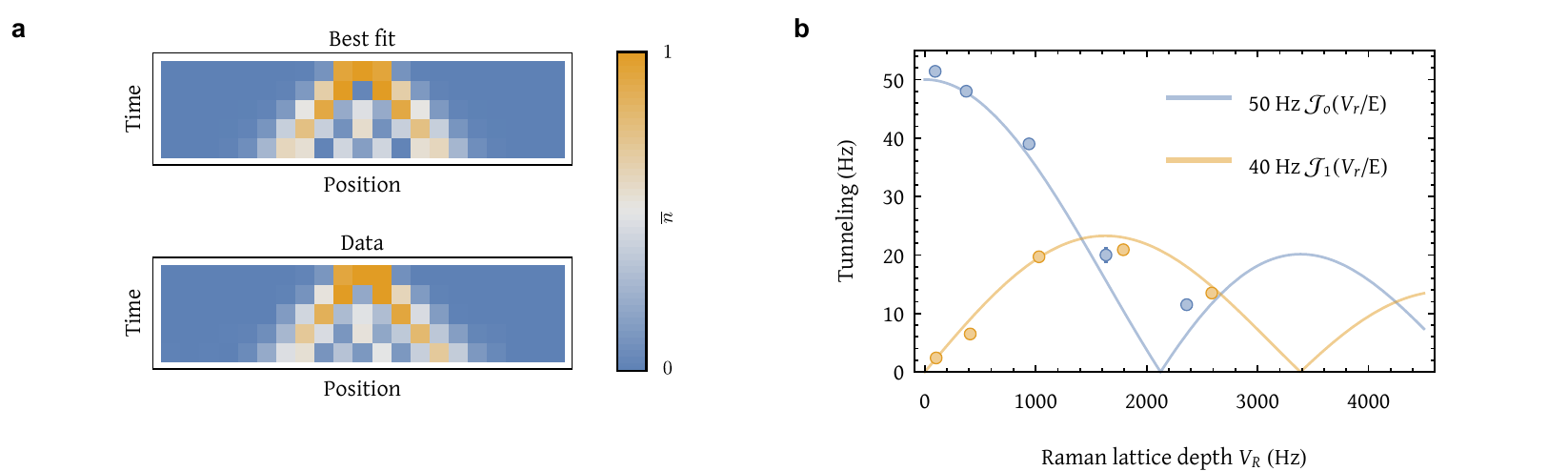}
    \caption{\label{fig:QWplot} \textbf{Restored and suppressed tunneling calibration}: \textbf{a}, Quantum walk in a one-dimensional lattice along the leg dimension $x$. \textbf{b}, Dependence of the tunneling rates along $x$ (yellow) and $y$ (blue) on the gauge field power. Here, $\mathcal{J}_{0,1}$ are Bessel functions of the first kind. They are scaled by the bare tunneling strengths 40\,Hz along $x$ and 50\,Hz along $y$ that we measure in the absence of tilt and Raman lattice.}
\end{figure*}

%Tunnelings calibration
The tunneling $J$ along the rung is determined from a double-well oscillation.
We initialize an atom on one side of a double-well potential and measure its time evolution.
The occupation number undergoes oscillations between the sites with frequency $2 J$. We obtain the tunneling rate $J/h = 34\,\text{Hz}$ from a fit to the data. In order to characterize the tunneling rate $K$ along the leg, we perform a single-particle quantum walk \cite{preiss_strongly_2015}. A single atom is initialized in a one-dimensional lattice along $x$ and the evolution of the density distribution is measured (Fig.~\ref{fig:QWplot} (a)). For a quantum walk, we expect the density distribution on site $i$ to evolve in time as
\[\rho_i \left( t \right) = \left| \mathcal{J}_i \left( \frac{2}{\pi} \frac{K}{\delta} \sin \left( \pi \delta t \right) \right) \right|^2,\]
where $\mathcal{J}_i$ is the Bessel function of the first kind on site $i$ and $\delta$ is the residual lattice tilt (equivalent to residual Raman lattice detuning)\cite{Hartmann2004}.
The theoretical density distribution is fitted to the data (Fig.~\ref{fig:QWplot} (a)), yielding $K/h = 11.4\,\text{Hz}$.
We also investigate the dependence of the tunneling rates on the power of the Raman beams by performing single-particle quantum walks along $x$ and $y$ for several Raman lattice depths.
The resulting tunneling rates and fits of the expected dependence are shown in Fig.~\ref{fig:QWplot} \cite{aidelsburger_measuring_2015}.

\subsection{Interaction and Tilt Calibration}
% On-site interaction and lattice tilt calibration
We use photon-assisted tunneling to calibrate both the on-site interaction $U$ and the lattice tilt $E$~\cite{Ma2011}.
We start with a unity-filling Mott insulator. By modulating the optical lattice potential, we observe photon-assisted tunneling between neighboring sites at the modulation frequencies $U \pm E$. From this, we extract $U/h = 131\,\text{Hz}$ after scaling to the lattice depth that was used in the experiment. In addition, we independently calibrate the dependence of the tilt $E$ on the voltage applied to the gradient coil.

\section{Experimental Sequence}
\begin{figure*}[!htbp]
    \centering
    \includegraphics{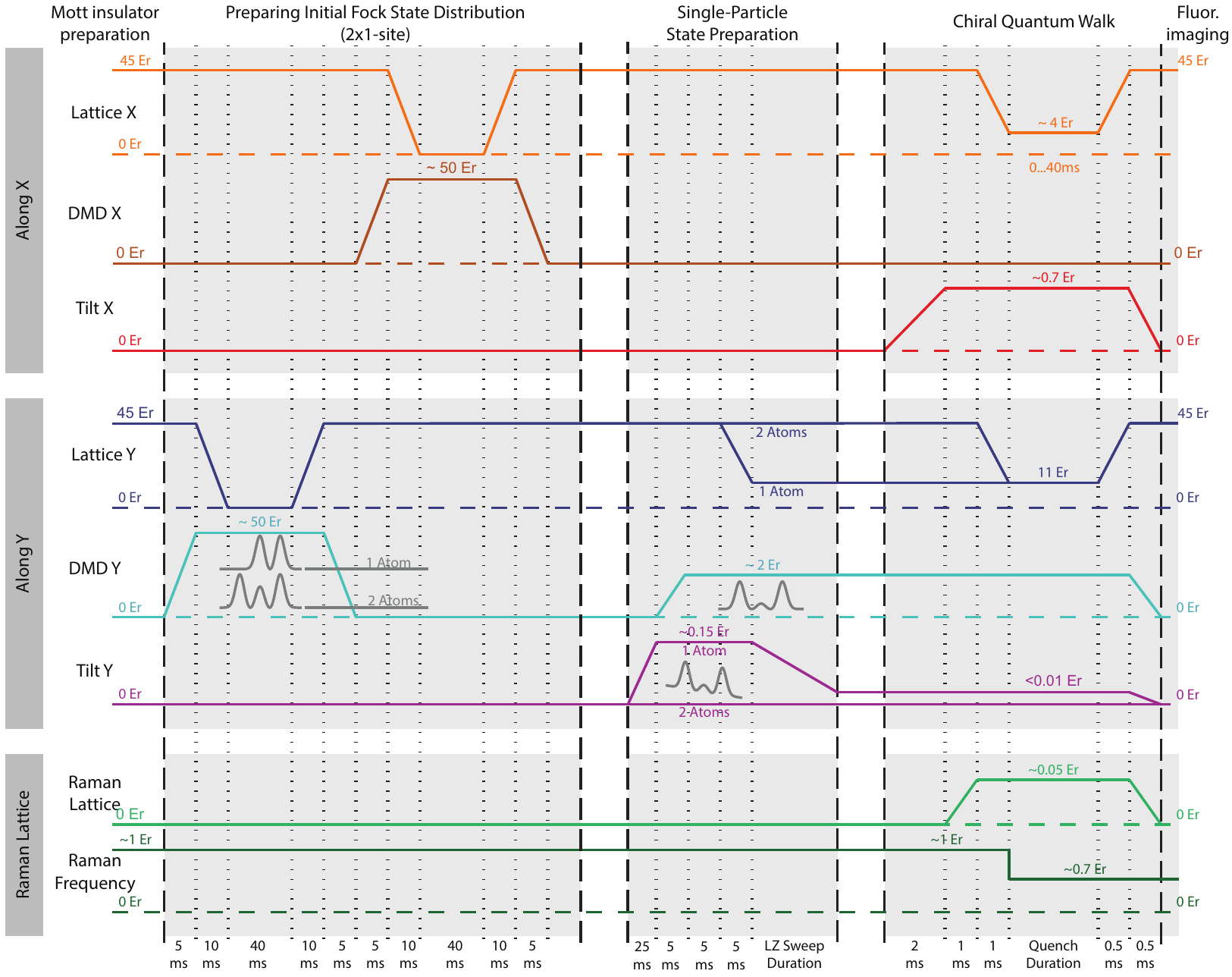}
    \caption{\label{fig:ExpSequencePlot} \textbf{Experimental sequence}: Schematic showing the approximate ramps and relative timing of the {\textit{x}}-, {\textit{y}}-lattices, the {\textit{x}}-, {\textit{y}}-DMD-potentials, the {\textit{x}}-, {\textit{y}}-tilts, and the Raman lattice. The profiles of the DMD potentials are sketched for the dimension labeled on the left. The other dimension of the profiles is well described by a smooth flattop potential within the region used for the experiment. All optical ramps are changed exponentially in depth as a function of time and are sketched with a logarithmic {\textit{y}}-scale here. The {\textit{x}}-, {\textit{y}}-tilts that are created by magnetic field gradients are, however, plotted and ramped linearly.}
\end{figure*}

All of the experiments described in this letter start with a 2-D, single layer Mott insulator of ${}^{87}$Rb in a deep optical lattice ($V_x=V_y = 45E_r$ , $E_r/2\pi\sim1.24 \, \text{kHz}$) with a $680 \, \text{nm}$ spacing~\cite{bakr_quantum_2009}. The experimental sequences for both the single-particle and interacting two-particle experiments are illustrated in Fig.~\ref{fig:ExpSequencePlot}.

\subsection{Initial State Preparation}

\subsubsection{Preparation of the Initial Fock State Distribution}

For both experiments we deterministically prepare an initial state from the $N=1$ shell of a Mott insulator. We choose the number of atoms for each experiment by projecting an additional confining (or ``cutting") optical potential from a digital micromirror device (DMD) located in the Fourier plane of our imaging system~\cite{zupancic_ultra-precise_2016}. This ``cutting" potential is either a single-well or a double-well along one dimension, and a smoothed flattop potential along the other dimension. The additional potential is superimposed on top of the atoms which are still situated in a deep lattice. The atoms outside of the ``cutting" potential are then removed from the system by turning off the optical lattice and applying an anti-confining potential to efficiently expel them from the system. The lattice is then ramped back on and the anti-confinement potential is turned off. As illustrated in Fig.~\ref{fig:InitStatePlot}, this sequence is first applied in the {\textit{x}}- and then in the {\textit{y}}-direction of the lattice such that either a $1 \times 1$ or a $2 \times 1$ initial state is produced. The loading efficiency of these states is $\approx93\%$ and is largely dominated by the initial Mott insulator fidelity.

\subsubsection{Single-Particle State Preparation}
\begin{figure*}[!htbp]
    \centering
    \includegraphics{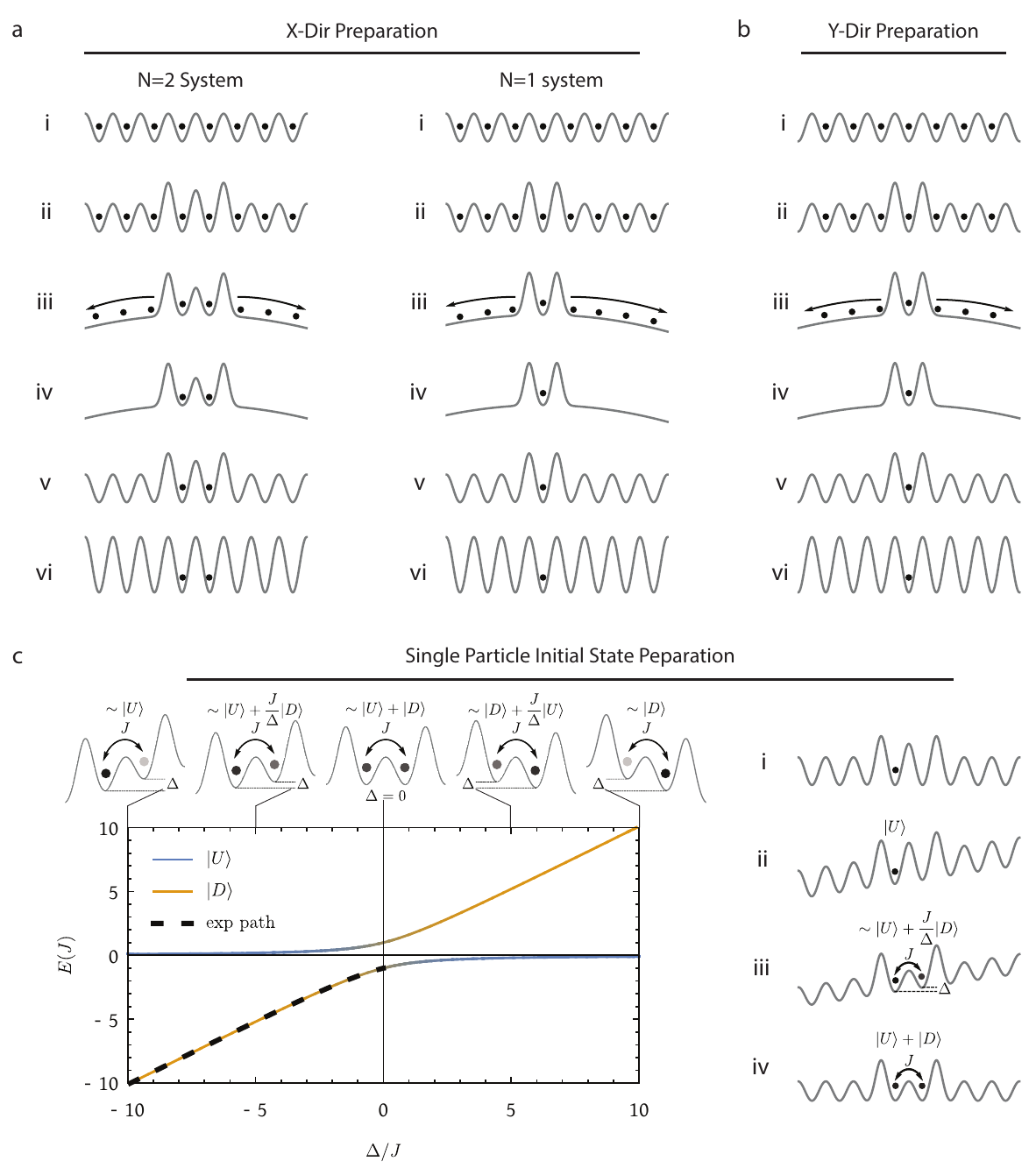}
    \caption{\label{fig:InitStatePlot} \textbf{Initial state preparation}: The steps for preparing the one- or two-particle states are shown in part \textbf{a} and \textbf{b} for the \textit{x}- and \textit{y}-directions, respectively. \textbf{i,ii} illustrate the superposition of the confining DMD potential and the bare lattice potential. \textbf{iii,iv} illustrate the removal of unwanted atoms from the system due to the removal of the optical lattice and the addition of an anti-confining potential. \textbf{v,vi} illustrate the reloading of the optical lattice and the removal of the additional ``cutting" potential from the DMD. The preparation of the delocalized single-particle initial state is illustrated in panel \textbf{c}. The left plot shows the energy of the two instantaneous eigenstates of the system as a function of $\frac{\Delta}{J}$ where the color denotes the overlap of the bare states $|U\rangle$ and $|D\rangle$ with these eigenstates. The initial state for the single-particle experiments is obtained by starting at $\frac{\Delta}{J}\approx20$ and adiabatically following the ground state to the $\Delta=0$ point in the diagram, as denoted by the dashed black line in panel \textbf{b}. The sketches above the plot visualize the approximate description of the system at given ratios of $\Delta/J$. The overall preparation sequence is visualized in the additioanl sketches \textbf{i-iv}. \textbf{i,ii} show the tilting of the lattice while tunneling is still strongly suppressed. \textbf{iii} shows the restoration of tunneling after the depth of the tilted lattice has been reduced. \textbf{iv} shows the final state produced by the Landau-Zener sweep after the tilt has been ramped down. The overall sequence thus corresponds to the state transformation $|U\rangle \rightarrow \frac{1}{\sqrt{2}}(|U\rangle + |D\rangle)$.}
\end{figure*}

There is an additional Landau-Zener preparation step for the single-particle experiments, which require an individual atom to be delocalized across the central rung of the ladder system. First, the ladder-forming double-well potential is projected on top of a deep, non-tilted lattice with a single atom being located on one side of the central double-well. While the tunneling is still suppressed, we add an additional tilt $\Delta\gg J$ with a physical magnetic field gradient along the rung direction, which is used to prepare the occupied site as the ground state of the tilted double-well system (cf. Fig.~\ref{fig:InitStatePlot} (c)). Tunneling is then rapidly increased by ramping down the lattice potential (final parameters: $V_y=11E_{r}$,$\frac{\Delta}{J}\approx 20$). Finally, we adiabatically prepare the $\Delta=0$ ground state of the balanced double-well system by ramping down the magnetic field gradient over a period of 100ms, which results in the state $|\psi_{\rm initial}\rangle= \frac{1}{\sqrt{2}}(a^{\dagger}_U + a^{\dagger}_D)|\text{vac}\rangle$. In doing so, we bring the gradient close to zero but eventually set it to a small finite value for the remainder of the experiment. This empirically chosen value compensates for any other sources of tilt in the double-well system and maintains the balanced population in the $|U\rangle$ and $|D\rangle$ states until the experiment is started.

\subsection{Quench Dynamics with a Gauge Field}
After the preparation of the initial state (either one or two particles), the dynamics are initiated by a quench in both directions. In the two-particle case, a double-well potential whose minima align with those of the \textit{y}-lattice is superimposed on the lattice to create a 2 $\times$ N confining potential. Afterwards, complex tunneling along the \textit{x}-direction is engineered by ramping on a physical magnetic field gradient while still maintaining a deep optical lattice. This gradient is large enough to suppress tunneling even after the next step, during which the optical lattice is ramped down to a lower depth. It is noteworthy that the resonance frequency for restoring the tunneling is calibrated at this lower lattice depth, since optical potentials of different strengths can have residual gradients resulting in a lattice-depth dependent resonance condition. To restore the tunneling suppressed by the tilt, the Raman lattice power is first ramped up to $V_R \approx 0.05 E_r$, then the optical lattice along the \textit{x}-direction is ramped down to $\approx 4 E_r$ to increase the bare tunneling in this direction. Afterwards, the frequency of the Raman lattice is chirped to the resonant frequency to enable complex tunneling and thereby realize the Harper-Hofstadter Hamiltonian on the 2 $\times$ N ladder that is defined by the bare lattice and the remaining DMD potential. In the two-particle case, the \textit{y}-lattice is additionally quenched to $11E_r$ at the time the Raman lattice is chirped to resonance. In the single-particle case, this is unnecessary since the lattice is left at $11 E_r$ from the Landau-Zener sweep. After a given quench time all lattices are ramped to their maximum depth to suppress all dynamics and the atoms are then imaged with single-site resolution (\ref{fig:ExpSequencePlot}). For all two particle experiments conducted, the doublon fraction at any time during the evolution is sufficiently small such that the results are not affected by the loss of atoms due to parity projection.

\section{Single-Particle Initial State Fidelity}
\begin{figure}[!htbp]
    \centering
    \includegraphics{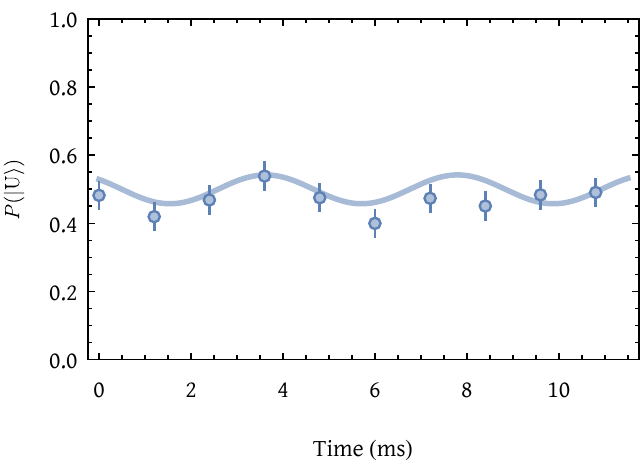}
    \caption{\label{fig:single_part_fid} \textbf{Preparation of delocalized single-particle states}: Probability to find the particle on the upper well over time. The curve describes the evolution of the fitted initial state ${|\psi_{\rm initial}\rangle = \sin{(\theta)}|U\rangle + \cos{(\theta)}e^{i \phi}|D\rangle}$ with $\theta/\pi=0.24(1)$ and $\phi/\pi=-0.02(2)$.}
\end{figure}

The initial state for the single-particle experiments would ideally correspond to ${|\psi_\text{ initial}\rangle=\frac{1}{\sqrt{2}}(|U\rangle+|D\rangle)}$. However, errors in the adiabaticity of the Landau-Zener ramp can result in either a population imbalance between the two wells or an additional relative phase. Generally then, this state can be written as ${|\psi_\text{initial}\rangle = \sin{(\theta)}|U\rangle + \cos{(\theta)}e^{i \phi}|D\rangle}$ where the desired state has $\theta/\pi=0.25,\phi/\pi=0$. To measure the population and phase imbalance of the initial state, we let the initial state evolve under just the rung dynamics at a lattice depth of $V_y\approx 6 E_r,V_x\approx 45 E_r$. The other lattice is at a depth of $V_x\approx 4E_r$, but tunneling in this direction is suppressed by the large tilt. The Raman beams that are used in the experiment to restore this tunneling are left off for this measurement. If both the amplitude and the phase were correctly balanced, then the evolution would show no dynamics between the sides of the initial rung and would have a single-well occupancy of $50\%$. The fitted evolution is shown in Fig.~\ref{fig:single_part_fid} and yields $\theta/\pi=0.24(1),\phi/\pi=-0.02(2)$. These values are almost equal to the ideal values mentioned above and provide a good calibration of the initial state preparation for the single-particle experiments where an additional phase shift in $\phi$ can arise from the rapid turn-on of the Raman beams.

\section{Single-Particle Dynamics along the x Direction}
\begin{figure}[!htbp]
    \centering
    \includegraphics{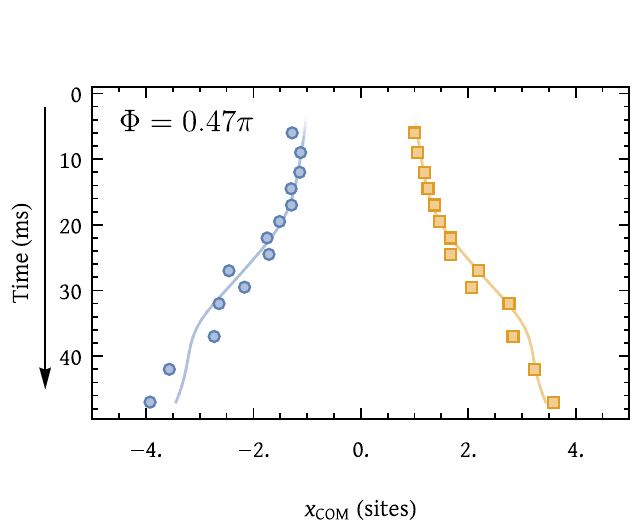}
    \caption{\label{fig:xcom_dynamics} \textbf{Single-particle dynamics along the x direction}: We separately determine the center-of-mass coordinates of the single-particle wavefunction for each half of the ladder (cf. figure 3 of the paper). The plot shows the x-coordinate of this center-of-mass vs. time.}
\end{figure}

The chiral nature of the single-particle dynamics is evident from the particle's center-of-mass motion along the rung direction, plotted separately for each ladder half in figure 3 of this paper. Figure \ref{fig:xcom_dynamics} shows the complementary plot for the x direction, describing the expansion of the single-particle wave function along the ladder. The data shows excellent agreement with theory for both ladder halves, implying a low amount of disorder in the system.

\section{Flux Calibration}

The flux per plaquette depends on the k-vector of the Raman lattice that is used to establish the laser-assisted tunneling~\cite{miyake_realizing_2013,aidelsburger_realization_2013}. The phase along each complex tunneling element is given by the dot product of the running wave formed by the two Raman beams and the displacement vector of the bare lattice, i.e. $\phi_{m,n}=\delta \textbf{k} \cdot \textbf{R}_{m,n}=m\phi_x + n\phi_y$ where $\phi_x = \delta \mathbf k \cdot a\mathbf e_x$ and $\phi_y = \delta \mathbf k \cdot a \mathbf e_y$ for our setup. Taking the line integral around a closed loop, one finds that this realizes a flux of $\Phi=\phi_y$ per plaquette, which is independent of $\phi_x$. The experiments described in this paper, however, were all performed with $\phi_x=\pi$ because this choice constitutes a wave vector along the \textit{x}-direction that maximally restores tunneling for a given Raman lattice depth. Considering the above-mentioned relations, the uniqueness of the presented apparatus lies in the tunability of the Raman lattice's $\mathbf{k}$ which directly implies the tunability of $\Phi$. The Raman lattice itself is produced by the interference of two blue-detuned optical beams that are projected through the same objective as the beams that produce the bare optical lattice. The $\mathbf{k}$ of the Raman lattice is tuned by two piezo mirrors which independently control the angle of the beams in the image plane. By independently changing the vectors $\mathbf{k}_1$ and $\mathbf{k}_2$, the $\delta \mathbf{k}$, and hence the flux $\Phi$, is easily tunable within an experimental sequence, which is not the case in other implementations that rely on a fixed angle and wavelength to produce a particular flux.

\begin{figure}[!htbp]
    \centering
    \includegraphics[width=0.9\columnwidth]{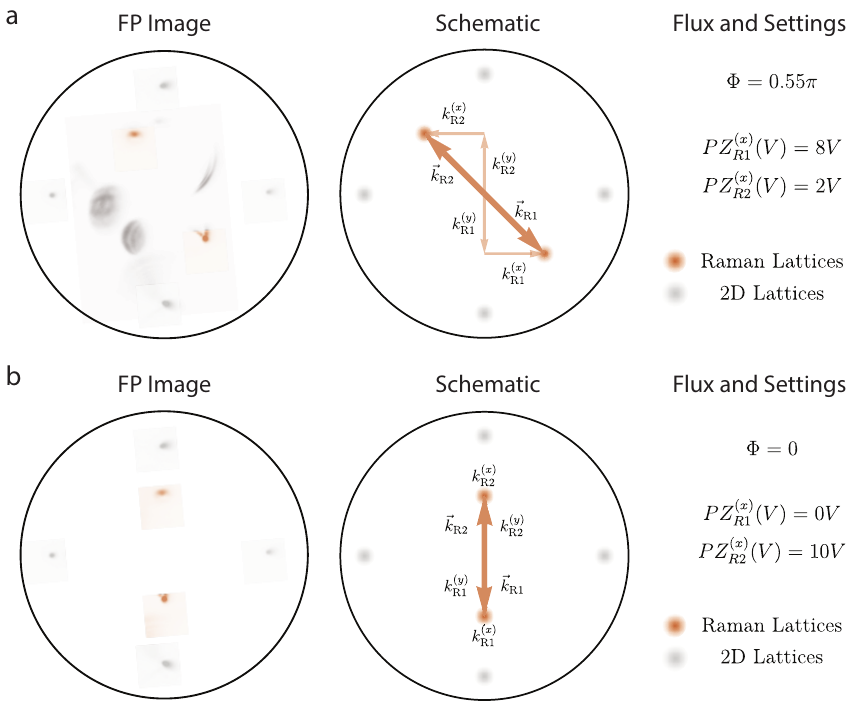}
    \caption{\label{fig:tuning_flux} \textbf{Tuning flux}: From left to right, the three columns show an image of the light fields in the Fourier plane (FP), a schematic visualizing the lattice vectors related to the Raman beams, as well as the voltages applied to the piezo mirrors to achieve the given configuration of these beams, both for $\Phi = 0.55\pi$ (\textbf{a}) and $\Phi = 0$ (\textbf{b}). The Fourier plane images are color coded as follows: The gray beams are the 2D lattice beams used to create the static optical lattice, the brown beams create the Raman lattice, and diffuse gray light represents unintended reflections off other optical surfaces.}
\end{figure}

\section{Heating rates in a Raman lattice}
\begin{figure*}[!htbp]
    \centering
    \includegraphics{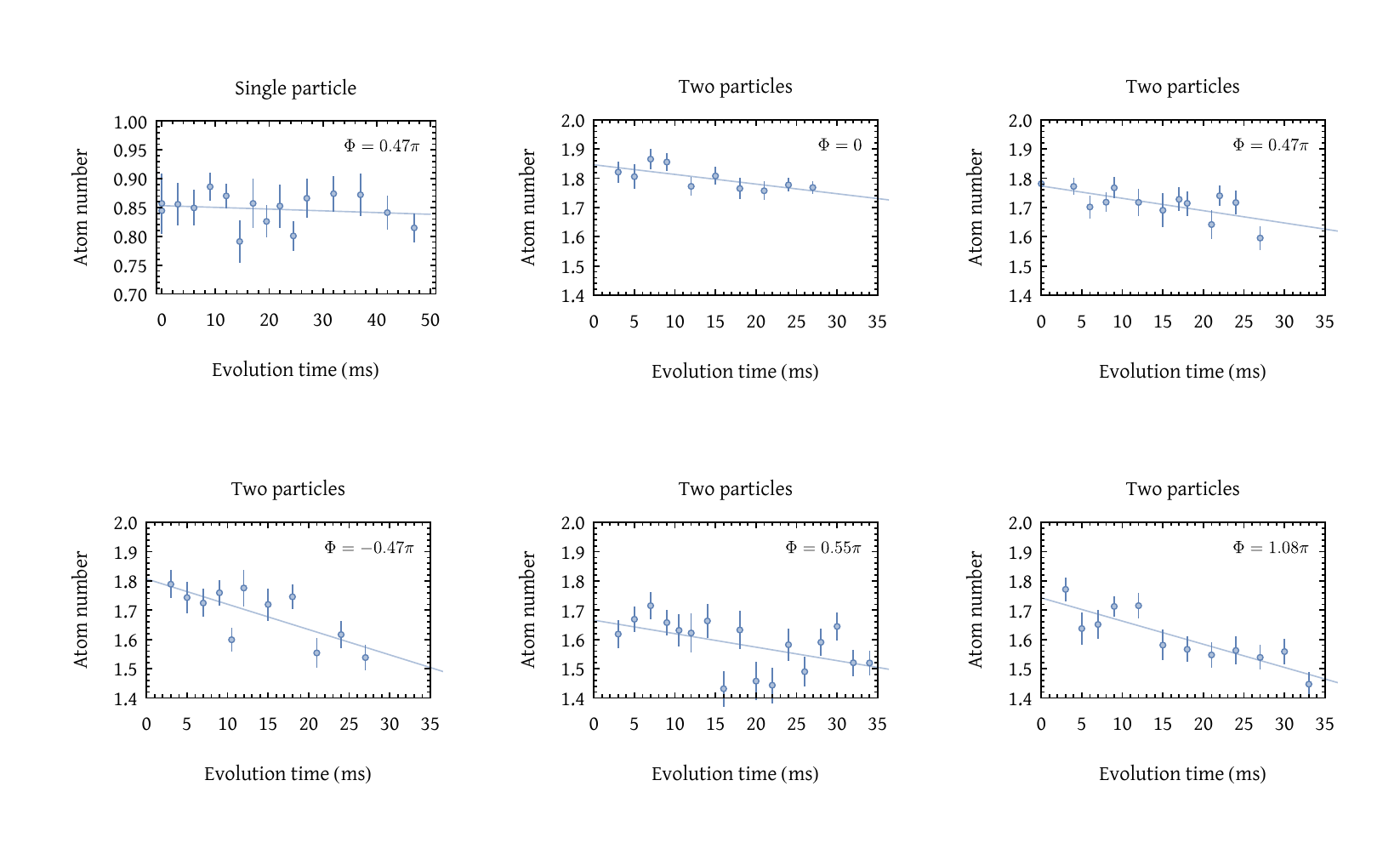}
    \caption{\label{fig:AtomNumberDecay} \textbf{Individual decay curves}: The figure shows the average atom number plotted against time for six experimental sequences realizing different flux values $\Phi$. For each sequence, a linear function is fitted to the data and the extracted slope is interpreted as the first order expansion of the exponential atom number decay in the system.}
\end{figure*}

We estimate heating rates in our Raman lattice by measuring the atom number decay as a function of time for different fluxes $\Phi$. Our experiments are performed on a time scale where this exponential decay is entirely dominated by its linear term, yielding
\[N(t)=N_0 \exp\left(-\frac{t}{t_0}\right) \approx N_0 - \frac{N_0}{t_0} t ,\]
where $N_0$ and $t_0$ are the initial atom number and the $1/e$ decay time, respectively. Using these quantities as free parameters, we fit individual decay curves to our experimental data, the results of which are shown in Fig.~\ref{fig:AtomNumberDecay}. 

\begin{figure}
    \centering
    \includegraphics{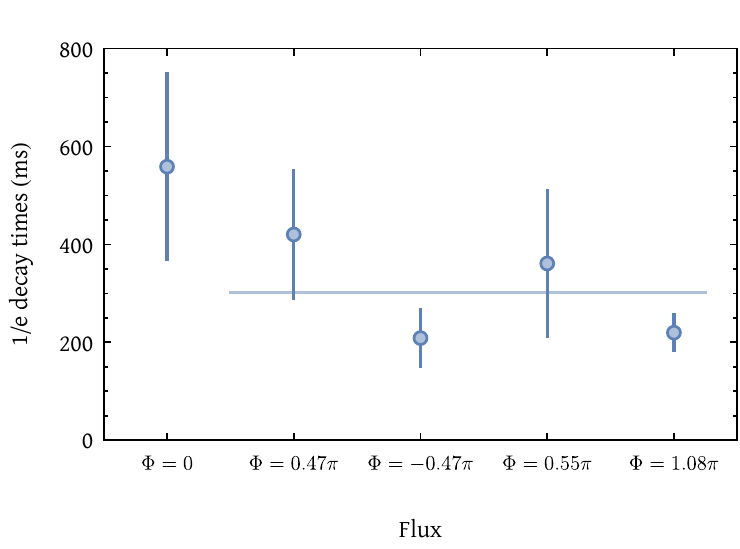}
    \caption{\textbf{Atom number decay rates}: The fitted decay rates for the five two-particle experiments plotted as a function of flux. Solid line is the average of decay rate for those cases where flux is non zero.}
    \label{fig:rates}
\end{figure}

Due to a variation of single-particle fidelities in the corresponding Mott insulators, the initial (average) atom number varies for different flux values $\Phi$, but the subsequent decay of the population is given by the heating dynamics in the Raman lattice. For the single-particle case with $\Phi = \pi/3$, the fitted decay time scale can not be distinguished from our vacuum-limited lifetime within the statistical error. For the experiments with two interacting particles in the ladder, we can compare the case without Raman lattice ($\Phi = 0$) to the case with photon-assisted tunneling for flux values $\Phi = \pi/3,-\pi/3,\pi/2$ and $\pi$ (cf. Fig.~\ref{fig:rates}). 

\end{document}